\documentclass[10pt,twocolumn,aps,pre,superscriptaddress]{revtex4-1} %

\usepackage{graphicx}
\usepackage{dcolumn}
\usepackage{amsmath}
\usepackage{amssymb}
\usepackage{bm}
\usepackage{color}
\usepackage{mathrsfs}
\usepackage{ulem}
\newcommand{\cref}[1]{(\ref{#1})}
\newcommand{\ud}{\mathrm{d}}

\begin{document}

\title{The Raspberry Model for Hydrodynamic Interactions Revisited. I.\protect\\ Periodic Arrays of Spheres and Dumbbells}

\author{Lukas P. Fischer}
\affiliation{Institute for Computational Physics (ICP), University of Stuttgart, Allmandring 3, 70569 Stuttgart, Germany}

\author{Toni Peter}
\affiliation{Institute for Computational Physics (ICP), University of Stuttgart, Allmandring 3, 70569 Stuttgart, Germany}

\author{Christian Holm}
\affiliation{Institute for Computational Physics (ICP), University of Stuttgart, Allmandring 3, 70569 Stuttgart, Germany}

\author{Joost de Graaf}
\email{jgraaf@icp.uni-stuttgart.de}
\affiliation{Institute for Computational Physics (ICP), University of Stuttgart, Allmandring 3, 70569 Stuttgart, Germany}

\date{\today}

\begin{abstract}
The so-called `raspberry' model refers to the hybrid lattice-Boltzmann and Langevin molecular dynamics scheme for simulating the dynamics of suspensions of colloidal particles, originally developed by [V. Lobaskin and B. D{\"u}nweg, New J. Phys. \textbf{6}, 54 (2004)], wherein discrete surface points are used to achieve fluid-particle coupling. This technique has been used in many simulation studies on the behavior of colloids. However, there are fundamental questions with regards to the use of this model. In this paper, we examine the accuracy with which the raspberry method is able to reproduce Stokes-level hydrodynamic interactions when compared to analytic expressions for solid spheres in simple-cubic crystals. To this end, we consider the quality of numerical experiments that are traditionally used to establish these properties and we discuss their shortcomings. We show that there is a discrepancy between the translational and rotational mobility reproduced by the simple raspberry model and present a way to numerically remedy this problem by adding internal coupling points. Finally, we examine a non-convex shape, namely a colloidal dumbbell, and show that the filled raspberry model replicates the desired hydrodynamic behavior in bulk for this more complicated shape. Our investigation is continued in [J. de Graaf, \textit{et al.}, J. Chem. Phys. \textbf{143}, 084108 (2015)], wherein we consider the raspberry model in the confining geometry of two parallel plates.
\end{abstract}

\maketitle

\section{\label{sec:intro}Introduction}

The physical description of hydrodynamic interactions in fluids has been a field of intensive study for over three centuries. The first mathematical description of (rarified) flow dates back to Euler.~\cite{euler57} This description was subsequently refined by Navier and Stokes to be applicable to the flow of dense media.~\cite{navier22,stokes49} However, finding solutions to the Navier-Stokes equations, even under the simplifying assumption of the low Reynolds number regime, has proven to be a particularly challenging boundary-value problem. Only in a few simple geometries can the Navier-Stokes equations be analytically solved, often leading to truncated series expansions rather than a full solution. 

Two geometries that can be handled semi-analytically are a simple-cubic array of spheres and a sphere between two parallel plates. The former is of particular interest as a toy model for fluid flow in a porous medium (at small sphere separations),~\cite{hofman99b} while the latter is relevant, for example, to the field of hydrodynamic chromatography.~\cite{giddings78,noel78} In this paper, we consider the crystalline arrangement and in Part II of our investigation,~\cite{degraaf15} we study the confining geometry of two parallel plates.

There are a myriad of (semi-)analytic investigations for the simple-cubic geometry, which makes this geometry perfectly suited for benchmarking the quality of hydrodynamic solvers. For the translational movement of a simple-cubic crystal through a fluid, the first results were obtained by Hasimoto, who derived a semi-numerical result for dilute systems.~\cite{hasimoto59} A complete numerical study for a larger range of lattice spacings and various crystal structures was later presented by Zick and Homsy.~\cite{zick82} The hydrodynamic flow around an infinite (simple-cubic) array of rotating spheres was first described by Brenner~\textit{et al.}~\cite{brenner70} These results were subsequently refined by Zuzovsky~\textit{et al.}~\cite{zuzovsky83} A complete numerical study of both translational and rotational friction over a large range of possible lattice spacings was provided by Hofman~\textit{et al.}~\cite{hofman99b} We utilize this large body of data as a reference throughout our manuscript.

A breakthrough in the numerical simulation of fluid dynamics resulted from the development of the lattice-Boltzmann (LB) algorithm. LB is based on the discretized version of the Boltzmann transport equation, see, \textit{e.g.}, Ref.~\cite{dunweg09} for a brief background. This lattice-based algorithm allows for the efficient simulation of hydrodynamic interactions in arbitrary geometries using simple boundary conditions, such as the bounce-back rule to obtain no-slip surfaces.~\cite{dunweg09} 

\begin{figure}[!htb]
\begin{center}
\includegraphics[scale=0.75]{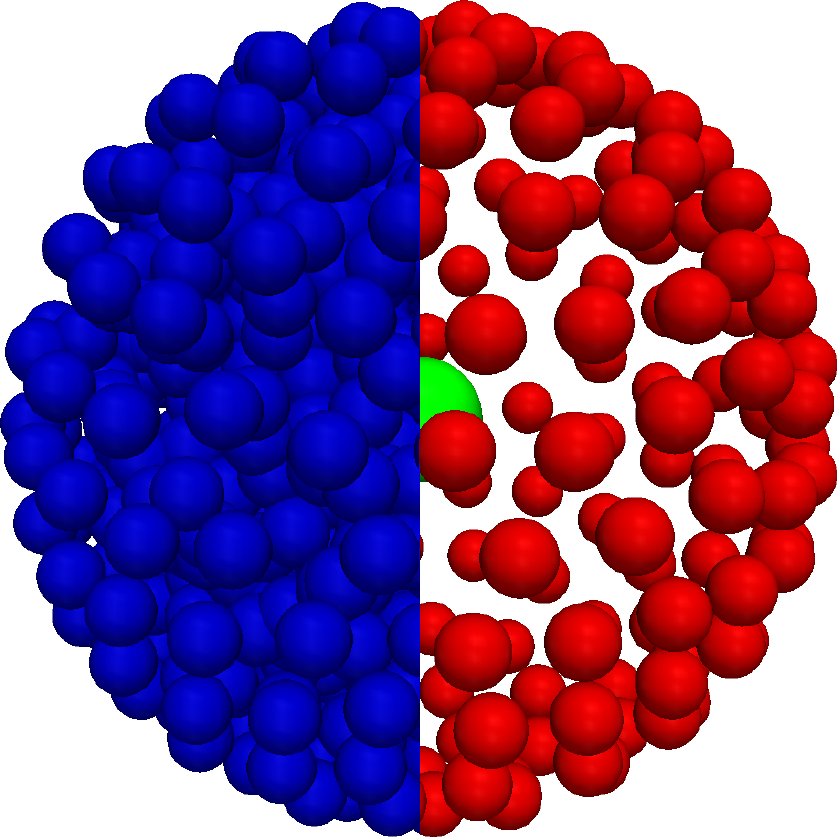}
\end{center}
\caption{\label{fig:rasp}(color online) Representation of the structure of the raspberry models used in our simulations, filled (left) and hollow (right), respectively. The central bead to which all other beads are connected \textit{via} rigid bonds, is shown using a green sphere. The blue spheres represent the beads that form the filled raspberry and the red ones give the surface beads used for the hollow variant. The radius of the beads is chosen to be smaller than the typical effective hydrodynamic radius to help visualize the internal structure. This figure is also included in Part II~\cite{degraaf15} of our analysis of the raspberry model.}
\end{figure}

One method to model particles moving in an LB fluid was introduced by Ahlrichs and D{\"u}nweg, who simulated polymer chains by utilizing an interpolated point-coupling scheme.~\cite{ahlrichs99} These points couple to the fluid through a frictional force, acting both on the solvent and on the solute, which depends on the relative velocity. The effect of this coupling is the formation of a hydrodynamic hull around the points, which thus gain a finite hydrodynamic extent (effective hydrodynamic radius).~\cite{ahlrichs99} Even if individual friction coefficients, and thus different effective radii, are used for the points, this method is limited in the effective size ratios that it can handle. Namely, by the particle-grid interpolation scheme and discretization used for the LB fluid.~\cite{ahlrichs99} Thus the method cannot be employed to study systems with substantial variation in particle size, for example, the electrophoresis of colloids with explicit ions. 

Lobaskin and D{\"u}nweg remedied this issue by introducing the so-called `raspberry' model, in which a larger colloid is modeled using the aforementioned coupling by discretizing the surface of the colloid into points.~\cite{lobaskin04} The method derives its name from this discretized nature of the surface, which resembles a raspberry, when represented by molecular-dynamics (MD) beads, see Fig.~\ref{fig:rasp}. A proper coverage of the surface by coupling points, such that the fluid inside of the shell is `trapped' and thus translates and rotates in unison with the shell, was assumed to create an effective no-slip/co-moving boundary condition at the surface.~\cite{lobaskin04,chatterji05}

Other methods to simulate moving boundaries exist, both LB-based and non-LB. Moving Ladd bounce-back (Ladd BB) boundaries exploit the lattice structure of the LB in describing colloidal particles.~\cite{ladd94,aidun95} The immersed boundary method (IBM),~\cite{Peskin02} and the external boundary force (EBF) method~\cite{Wu10} both use a point-coupling strategy to describe particles in LB; although it has recently been shown that these methods are special choices of the friction and mass ratio in the Ahlrichs and D{\"u}nweg scheme.~\cite{schiller14} The most commonly used non-LB methods include: dissipative particle dynamics (DPD),~\cite{Hoogerbrugge92,Espanol95} multi-particle collision dynamics (MPCD) or stochastic rotation dynamics (SRD),~\cite{Malevanets99,Ihle03} and Stokesian Dynamics (SD).~\cite{brady88} However, the raspberry method has remained popular, because of its simplicity as a straightforward extension of point-particle coupling. It has been extended upon~\cite{chatterji05} and has been used in a wide variety of simulation settings.~\cite{lobaskin07,lobaskin08,raafatnia14} Recently, this model was employed in the context of multi-particle collision dynamics (MPCD),~\cite{belushkin11,poblete14} stochastic rotation dynamics (SRD),~\cite{sane09} and dissipative particle dynamics (DPD) simulations.~\cite{lugli11,zhou13} 

Of singular interest are a set of recent publications from the Denniston group.~\cite{Ollila12,ollila13,Mackay13a,Mackay13b} In these publications the quality of the (raspberry-type) point-coupling schemes are investigated and compared to theoretical expressions. Ollila~\textit{et al.} show in Ref.~\cite{Ollila12} that there is good correspondence between the LB simulations and analytic results~\cite{Deutch75,Felderhof75b} for a hollow shell, an annulus, and a dense distribution of coupling points. They place these results in the context of the simulation of porous particles. In Ref.~\cite{ollila13}, Ollila~\textit{et al.} further analyzed the quality of the point-coupling method and showed that there are problems with this scheme when utilizing it to describe solid particles. In particular, Ollila~\textit{et al.} demonstrated that the hydrodynamic radius of these particles is ill-defined in an LB fluid. That is, the effective hydrodynamic radius that follows from the translational mobility (\textit{via} the Stokes relation) does not match that obtained using the rotational mobility. By careful calibration,~\cite{ollila13} the use of a colloid radius that is `incommensurate' with the lattice spacing,~\cite{Ollila12} and modification of the coupling of the points to the LB fluid,~\cite{Mackay13a,Mackay13b} the rotational and translational effective radii can be well-matched in the formalisms suggested by Ollila~\textit{et al.} and coincide with the radius given to the coupling points.

In this manuscript, we re-examine the raspberry model by Lobaskin and D{\"u}nweg~\cite{lobaskin04} in the context of the work of Ollila~\textit{et al.}~\cite{Ollila12,ollila13} We show that there is a simple way to obtain an effectively consistent hydrodynamic description of a solid particle using the raspberry model for suitably chosen LB and coupling parameters. Namely, by a `filling + fitting' strategy, which we will describe in detail in Section~\cref{subb:FSS}. This approach consists of introducing coupling points to the interior of the raspberry particle and fitting for the radius of a solid particle using suitable experiments. Our `filling + fitting' procedure does not necessitate a particle radius that is incommensurate with the lattice. Moreover, it yields an internally consistent formalism, which reproduces the hydrodynamic properties of a solid object with a high degree of accuracy. 

We show how our fit parameter (the effective hydrodynamic radius) can be straightforwardly determined. To demonstrate that our method works for a range of reasonable LB parameters, we examine the quality of the raspberry model in the classic fluid-dynamics geometry of a simple-cubic arrangement.~\cite{hasimoto59,brenner70,zuzovsky83,hofman99b} We show that the raspberry model reproduces the theoretical result surprisingly well over the complete range of applicable raspberry (sphere) separations. In obtaining these results, we also analyze the quality of the standard hydrodynamics experiments performed in this geometry.~\cite{lobaskin04,chatterji05} We further demonstrate that the improved correspondence between the effective rotational and translational hydrodynamic radius is upheld over a large range in bare frictions, (original/imposed) radii of raspberry, and sufficiently large filling fractions. We also comment on the interpretation of our data in the context of theoretical results for porous objects.~\cite{Debye48,Felderhof75a,Felderhof75b} Finally, we consider the effectiveness of the raspberry description in modeling solid non-convex particles and show that the `filling + fitting' model gives accurate results for the bulk mobility of a dumbbell-shaped colloid. Part II of our analysis is presented in Ref.~\cite{degraaf15} and extends these conclusions to raspberry particles under confinement. We thus demonstrate that for a wide range of suitably chosen parameters our `filling + fitting' formalism leads to a substantially improved (and acceptable) numerical tolerance in simulating solid objects with respect to that of the traditional raspberry model of Refs.~\cite{lobaskin04,chatterji05}.

The remainder of this manuscript is structured as follows. In Section~\ref{sec:methods} we describe our simulation methods in detail. Section~\ref{sub:LBmeth} opens with a description of the LB and coupling method. Section~\ref{sub:rasp} introduces our variant of the raspberry model for the spherical and dumbbell-shaped colloids of interest. Sections~\ref{sub:MD_para}~and~\ref{sub:LB_para} detail the molecular dynamics and LB simulation parameters, respectively. Section~\ref{sub:exper} describes the various hydrodynamic experiments that we performed to determine the properties of the raspberry model. In Section~\ref{sub:dimnum} we discuss the dimensionless numbers that characterize the physics of our systems. We provide a summary of the notations used throughout the text in Section~\ref{sub:notation} to aid the reader when going through the manuscript. In Section~\ref{sec:result} we list our main results. We begin by examining the properties of the spherical raspberry in a simple-cubic lattice in Section~\ref{sub:SCsphere}. We continue with the properties of two dumbbell-shaped raspberries with different geometric parametrizations in Section~\ref{sub:SCdumb}. The results are discussed and related to previous studies in Section~\ref{sec:disc}. Finally, we give a summary, conclusions, and an outlook in Section~\ref{sec:conc}.

\section{\label{sec:methods}Methods}

In this section, we outline the approach used to determine the hydrodynamic properties of a colloid. We have split this into subsections detailing the properties of the lattice-Boltzmann method, the construction of the raspberry model, the molecular dynamics and lattice-Boltzmann parameters used, the hydrodynamic experiments performed to extract the mobility of the raspberry, the dimensionless numbers that characterize the fluid, and a reference list of the input parameters and measured quantities.

\subsection{\label{sub:LBmeth}The Lattice-Boltzmann Method}

In this section, we briefly outline the major features of the lattice-Boltzmann method and viscous particle coupling to put our work into context. We refer the interested reader to, for instance, Ref.~\cite{dunweg09} for a more in-depth treatment. 

The LB method is a numerical simulation technique to solve the Boltzmann transport equation.~\cite{boltzmann64} In its simplest form the Boltzmann equation can be written as
\begin{equation}
\label{eq:BT} \partial_{t} f(\mathbf{r},\mathbf{v},t) + \mathbf{v} \cdot \mathbf{\nabla} f(\mathbf{r},\mathbf{v},t) = C(f(\mathbf{r},\mathbf{v},t)),
\end{equation}
where $t$ denotes time, $\mathbf{r}$ the position, and $\mathbf{v}$ the velocity; $\partial_{t}$ indicates a partial derivative with respect to time, $\cdot$ indicates the dot product, and $\mathbf{\nabla}$ the gradient with respect to position; and $f(\mathbf{r},\mathbf{v},t)$ is a phase-space probability distribution function and $C(f(\mathbf{r},\mathbf{v},t))$ is the collision operator acting on the distribution function, which models the probability redistribution caused by particle interactions.

The lattice-Boltzmann equation is the discretized form of Eq.~\cref{eq:BT}, where the particle velocities are restricted to only a few values. The LB `particles' can thus only move in a finite number of directions, which are chosen to be commensurate with a space-filling lattice. When this lattice has sufficient symmetry to fulfill mass and momentum conservation, the discrete LB equation can be used to determine fluid flow, without directly solving the Stokes or Navier-Stokes equations, as has been shown via the Chapman-Enskog expansion.~\cite{chapman91} The physical quantities that are of interest, such as the mass density, velocity, and pressure, can be recovered from the modes of the discrete probability distribution.

Current implementations of the LB method trace their roots to the lattice gas automata that were developed in the late 1980s.~\cite{frisch86,wolfram86} The traditional LB method was formulated by making an assumption for the form of the collision operator, the right-hand side of Eq.~\cref{eq:BT}, the most well-known being the single-relaxation scheme introduced by Bhatnagar, Gross, and Krook.~\cite{Bhatnagar54} The LB method has significant advantages over traditionally used fluid solvers, as the algorithm is completely local, which allows for straightforward parallelization.~\cite{dunweg09} Moreover, the streaming operator (left-hand side of Eq.~\cref{eq:BT}) and the collision process can be fully decoupled, leading to an algorithm that is elegant in its simplicity.

The LB method can be connected to a Molecular Dynamics solver, in order to model the behavior of particles suspended in a viscous fluid. One method to achieve particle-fluid coupling was proposed by Ahlrichs and D{\"u}nweg.~\cite{ahlrichs99} The fluid is coupled to embedded MD beads via a friction force that depends on the difference in velocity between the bead and the fluid
\begin{equation}
\label{eq:couple} \mathbf{F}_{d} = - \zeta_{0} \left( \mathbf{u}_{p} - \mathbf{u}_{f}(\mathbf{r}_{p}) \right),
\end{equation}
where $\mathbf{F}_{d}$ is the friction force, $\zeta_{0}$ is the bare friction coefficient, $\mathbf{u}_{p}$ is the particle's velocity, and $\mathbf{u}_{f}$ is the fluid velocity that is evaluated at the particle's position $\mathbf{r}_{p}$. Here, the particle's coordinates are interpolated onto the lattice using a tri-linear scheme.~\cite{dunweg09} The opposite force has to be applied to the fluid to ensure momentum conservation. This algorithm is used to couple the beads of the raspberry model to the LB fluid that will be described in the next section.

\subsection{\label{sub:rasp}The Raspberry Model}

In this manuscript we study the so-called `raspberry' model for particle-fluid interactions,~\cite{lobaskin04} as shown in Fig.~\ref{fig:rasp}. This model relies on discretizing the surface of a larger colloid into coupling points, which experience a friction force related to the relative velocity of the fluid and the coupling points as described above.~\cite{ahlrichs99} In Ref.~\cite{lobaskin04}, 100 points were used to approximate a sphere. To ensure a reasonably homogeneous surface coverage these were connected to each other by finite extensible nonlinear elastic (FENE) potentials. The forces acting on the surface beads were forwarded to a central Lennard-Jones (LJ) MD bead, \textit{via} the LJ interaction. A model similar in spirit to the one proposed by Lobaskin and D{\"u}nweg was developed by Chatterji and Horbach.~\cite{chatterji05} In their construction the surface beads were fixed with rigid bonds to the central bead and no FENE potential was employed for the surface-surface coupling.

\subsubsection{\label{subb:hollow}The Hollow Raspberry}

For the construction of the raspberry model in this paper, we combined the approaches of Refs.~\cite{lobaskin04,chatterji05}. To homogeneously arrange the MD beads in a spherical shell of radius $R$, we used a separate MD simulation. We placed $N \gtrsim \lceil 4\pi R^{2}/a^{2} \rceil$ MD beads in a cubic simulation box with edge length $L$, LB lattice spacing $a$, and periodic boundary conditions. The number of MD beads was chosen such that on average there is at least one particle per lattice site for the LB simulation. To force the beads onto a spherical shell we employed a shifted harmonic bond potential around the center of the box, $\mathbf{r}_{P}$, which will become the center of the raspberry particle that we are creating. This potential has the form
\begin{equation}
\label{eq:shiftedHarm}V_{\mathrm{harm}}(\mathbf{r}) = \frac{1}{2}K\left(|\mathbf{r}-\mathbf{r}_{P}| - R \right)^{2} ,
\end{equation}
where $\mathbf{r}$ is a point in space and $K$ is the spring constant. To ensure that the beads do not overlap and to homogenize the surface density, we endowed them with a repulsive Weeks-Chandler-Anderson (WCA) interaction potential
\begin{equation}
\label{eq:WCA} V_{\mathrm{WCA}} = \begin{cases}
\displaystyle 4\epsilon\left(\left(\frac{\sigma}{r}\right)^{12} - \left(\frac{\sigma}{r}\right)^{6} + \frac{1}{4}\right) & r<2^{1/6}\sigma \\
0 & r \ge 2^{1/6}\sigma
\end{cases} ,
\end{equation}
where $\sigma$ is the MD base unit of length and is thus equal to the bead diameter. 

The MD beads were thermalized using a Langevin thermostat with `temperature' $1.0\epsilon$ and friction coefficient $\Gamma = 1.0 m_{0}\tau^{-1}$. Here, $\epsilon$ is the MD base unit of energy and corresponds to $1 k_{\mathrm{B}}T$, where $k_{\mathrm{B}}$ is the Boltzmann constant and $T$ is the temperature, $\tau$ is the MD base unit of time, and $m_{0}$ the MD base unit of mass ($m_{0} = \tau^{2}\epsilon\sigma^{-2}$). The MD beads were each given a mass $1 m_{0}$. By geometrically increasing the spring constant from $K = 1.0 \epsilon\sigma^{-2}$ to $K = $~3,000$\epsilon\sigma^{-2}$ the MD beads are forced onto the spherical shell described by the potential in Eq.~\cref{eq:shiftedHarm}. We increased $K$ to its final value of $K = $~3,000$\epsilon$ over 100,000 integration steps of length $\Delta t = 0.003\tau$. These simulations were performed using the MD software package \textit{ESPResSo}.~\cite{limbach06a,arnold13a} Finally, small deviations of the MD beads' radial position with respect to the desired distance $R$ were removed by adjusting their radial position. The configuration was then `frozen in' by connecting all beads to a central bead via rigid bonds (virtual sites).~\cite{arnold13a}

To test the quality of the result, the raspberry was checked for large holes in the surface coverage by applying a `shotgun' algorithm. We randomly picked 50,000 points on the surface of the sphere and calculate the distances to the nearest surface bead. We arrived at the distribution of MD beads that we used throughout our simulations, by repeating this procedure with different initial configurations and particle numbers, until we found a system for which the maximum hole size was roughly $1.0\sigma$ (bead diameter). The outcome for a sphere of radius $R = 3.0\sigma$ is shown in the right-hand side of Fig.~\ref{fig:rasp}. Here, 202 surface beads were used to obtain a maximal hole diameter of $1.1\sigma$. In total five variants of the hollow raspberry were considered, with radii $R = 2.0\sigma$, $2.5\sigma$, $3.0\sigma$, $4.0\sigma$, and $5.0\sigma$ and $N = 89$, $139$, $202$, $441$, and $593$, respectively. We also considered a `dense shell raspberry', with $R = 3.0\sigma$ and $N = 924$ beads in the shell, which will be discussed further in Section~\ref{subb:FSS}. Unless otherwise specified, whenever we use the term `hollow raspberry' in this document and Part II,~\cite{degraaf15} we refer to the raspberry with $N = 202$ surface beads and radius $R = 3.0\sigma$.

Finally, we should mention that there is an alternative method of positioning the coupling points on the shell. For the harmonic potential in Eq.~\cref{eq:shiftedHarm} and WCA interactions between the MD beads, a conjugate-gradient descent method can be used to generate a surface coverage with minimal defects.~\cite{Altschuler97}

\subsubsection{\label{subb:filled}Filling the Raspberry}

We `fill' the hollow-shell raspberry particle by adding coupling points to the interior, as outlined in detail below. We first formed a hollow raspberry according to the recipe in Section~\ref{subb:hollow}. Next, we added $N' \gtrsim \lceil 4 \pi (R - \sigma/2)^{3}/3 a^{3} \rceil$ beads to the interior of the shell, which interact with each other and the shell MD beads via the WCA potential of Eq.~\cref{eq:WCA}. The force between the internal beads themselves was initially capped to $1.0\epsilon/\sigma$ to prevent numerical instabilities. The system was allowed to evolve by making use of a Langevin thermostat ($k_{\mathrm{B}} T = 1.0\epsilon$, $\Gamma = 1.0 m_{0} \tau^{-1}$). The simulation consisted of over 50,000 time steps of length $\Delta t = 0.005 \tau$. During this time the capping value was slowly raised to $100\epsilon$. This generally resulted in a random configuration with a homogeneous distribution of MD beads within the raspberry. These beads were subsequently frozen in place by adding rigid (virtual) bonds to the central MD bead. 

We investigated several values of $N'$ and the role of the MD beads' distribution on the model's ability to reproduce the result of Stokes' equation, see Section~\ref{subb:porous}. We settled upon a value of $N' = 722$, resulting in a total of $N_{\mathrm{tot}} = N + N' + 1 = 925$ MD beads for the so-called `filled raspberry' of radius $R = 3.0\sigma$. This result is shown in the left-hand side of Fig.~\ref{fig:rasp}. Note that we used exactly the same hollow shell to construct our filled variant. For the raspberries with radius $R = 2.0\sigma$, $2.5\sigma$, $4.0\sigma$, and $5.0\sigma$, we used $N' =154$, $143$, $967$, and 1,323 internal coupling points, respectively. To study the improvement of the coupling on the internal filling factor, we also considered several other values of $N'$ for the $R = 3.0\sigma$ raspberry, namely: $N' = 50$, $100$, $200$, and $400$. Unless otherwise specified, we refer to the $R = 3.0\sigma$ and $N' = 722$ model as the `filled raspberry,' both here and in Part II.~\cite{degraaf15} 

Finally, it should be remarked that in the hydrodynamic simulations utilizing the raspberry model, all WCA interactions were switched off and only the rigid (virtual) bonds remained. This eliminated any non-hydrodynamic interactions between the raspberry and its images in our simulations with periodic boundary conditions for small box lengths ($L \approx 2R$).

\subsubsection{\label{subb:dumbbell}Constructing a Dumbbell Raspberry}

A dumbbell-shaped raspberry model (filled or hollow) is constructed using a procedure that is analogous to the one given in Sections~\ref{subb:hollow} and~\ref{subb:filled}. Instead of a central harmonic potential, we used two harmonic potentials centered on $\mathbf{r}_{P} = (0,0,- d/2)$ and $\mathbf{r}_{P}' = (0,0,d/2)$, with $d$ the distance between the sphere centers of the dumbbell (the total length of the dumbbell is $d + 2R$). In addition, a WCA potential had to be added to prevent beads from accumulating in the neck of the dumbbell -- the region where the two dumbbell spheres overlap, if $d < 2 R$. To accomplish this, we used a WCA potential between the center of the dumbbell, located at $(0,0,0)$, and the surface MD beads. This potential had the following form 
\begin{equation}
\label{eq:WCAneck} V_{\mathrm{neck}} = \begin{cases}
\displaystyle 4\epsilon\left(\left(\frac{w}{r}\right)^{12} - \left(\frac{w}{r}\right)^{6} + \frac{1}{4}\right) & r<2^{1/6}w \\
0 & r \ge 2^{1/6}w
\end{cases} ,
\end{equation}
where $w$ is the width of the neck and is given by 
\begin{equation}
\label{eq:neckwidth} w = \sqrt{R^2 - \frac{d^{2}}{4}}.
\end{equation}

\begin{figure}[!htb]
\begin{center}
\includegraphics[scale=1.0]{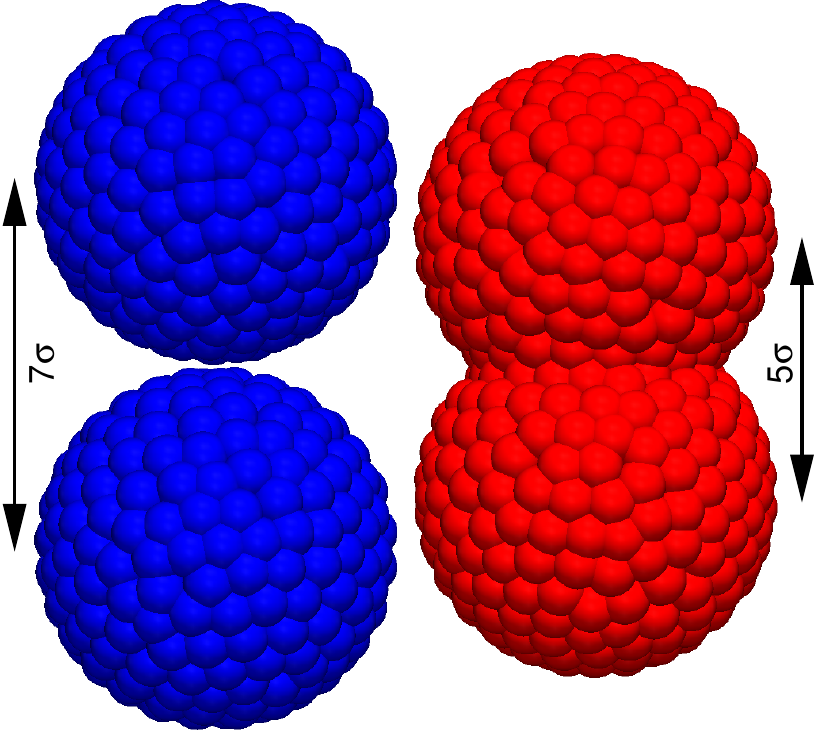}
\end{center}
\caption{\label{fig:dumb}(color online) Representation of the raspberry dumbbells used in our simulations, touching (left) and overlapping (right), respectively. The distance between the centers of the spheres (each $R=3.0\sigma$ in size) is indicated using the arrows. Note that we used the effective MD bead diameter of approximately $1\sigma$ to visualize our result.}
\end{figure}

After letting the MD beads become trapped in the dumbbell shell, in the same manner as for the spherical shell, they were connected \textit{via} rigid bonds to a particle at the geometric center of the dumbbell. The dumbbell may be filled with $N'$ additional beads using the procedure outlined in Section~\ref{subb:filled}. In this paper, we consider two dumbbell-shaped raspberry particles -- one with $d = 5.0\sigma$ and one with $d = 7.0\sigma$; for both the individual sphere radius is $R = 3.0\sigma$ -- corresponding to a partially overlapping configuration and one with the spheres just touching, respectively; see Fig.~\ref{fig:dumb}. We used ($N = 416$, $N' = 598$) for $d = 5.0\sigma$ and ($N = 502$, $N' = 404$) for $d = 7.0\sigma$, respectively, to ensure a homogeneous surface distribution and filling of the volume.

\subsection{\label{sub:MD_para}Molecular Dynamics Parameters}

Once we had constructed the raspberries, we could use them in our LB simulations. The raspberry particles were allowed to freely move and rotate, unless otherwise specified. All the forces acting on the MD beads are transferred to the central bead \textit{via} the virtual sites (rigid bonds). To stabilize the simulation for the bare friction coefficients used, we set the (bare) mass and rotational inertia of the raspberry; these quantities should not be confused with the virtual mass of the body in a fluid, see, \textit{e.g.}, Ref.~\cite{Zwanzig75} for the definition. The mass and rotational inertia are based on the particle's dimensions and the fluid mass density, which we denote by $\rho$ and set to $\rho = 1.0 m_{0}\sigma^{-3}$. We thus assume that the raspberry particle has the same density as the surrounding fluid.

For the spheres with radii $R = 2.0\sigma$, $2.5\sigma$, $3.0\sigma$, $4.0\sigma$, and $5.0\sigma$ the mass we used, was $m = (4/3) \pi \rho R^{3} \approx 33.5 m_{0}$, $65.5 m_{0}$, $113 m_{0}$, $268 m_{0}$, and $524 m_{0}$, respectively. The inertia tensor is a diagonal tensor with identical entries of $I = (8/15) \pi \rho R^{5} \approx 53.6 m_{0}\sigma^{2}$, $164 m_{0}\sigma^{2}$, $407 m_{0}\sigma^{2}$, $1.72\cdot10^{3} m_{0}\sigma^{2}$, and $5.23 \cdot 10^{3} m_{0}\sigma^{2}$ for these radii, respectively. For the two dumbbell raspberries, we used 
\begin{equation}
\label{eq:mdumb} 
m =\begin{cases}
\pi \rho \left( \displaystyle \frac{4}{3}R^{3} + dR^{2} - \frac{1}{12}d^{3}\right)  & 0 \le d < 2R \\[0.4cm]
\displaystyle \frac{8}{3}\pi \rho R^{3} & d \ge 2R
\end{cases} .
\end{equation} 
The dumbbell's rotational inertia tensor is diagonal, but the entries are not identical. Let $I_{\parallel}$ denote the moment for rotation about the main axis of the dumbbell and $I_{\perp}$ the moment for rotation about a central axis perpendicular to the main axis. We may then write
\begin{widetext}
\begin{eqnarray}
\label{eq:Iperpdumb} 
I_{\perp} & = & \begin{cases}
\pi \rho \left( \displaystyle \frac{8}{15}R^{5} + \frac{3}{4} dR^{4} + \frac{1}{3}d^{2}R^{3} + \frac{1}{24} d^{3}R^{2} + \frac{1}{960}d^{5} \right) & 0 \le d < 2R \\[0.4cm]
\pi \rho \left( \displaystyle \frac{8}{15} R^{5} + \frac{2}{3} d^{2}R^{3} \right) & d \ge 2R
\end{cases} ; \\
\label{eq:Iparadumb} 
I_{\parallel} & = & \begin{cases}
\pi \rho \left(\displaystyle  \frac{16}{15}R^{5} + \frac{1}{2} dR^{4} - \frac{1}{12}d^{3}R^{2} + \frac{1}{160} d^{5} \right)  & 0 \le d < 2R \\[0.4cm]
\displaystyle \frac{16}{15} \pi \rho R^{5} & d \ge 2R
\end{cases} ; \\
\label{eq:Idumb}
\uuline{I} & = & \left( \begin{array}{ccc}
I_{\perp} & 0 & 0 \\
0 & I_{\perp} & 0 \\
0 & 0 & I_{\parallel} \end{array} \right),
\end{eqnarray}
\end{widetext}
where the long axis is assumed to be aligned with the $z$-axis. This gives us the following for the $d = 5.0\sigma$ dumbbell: $m \approx 221m_{0}$, $I_{\perp} \approx 2.23\cdot10^{3} m_{0}\sigma^{2}$, and $I_{\parallel} \approx 810m_{0}\sigma^{2}$. Whereas for the $d = 7.0\sigma$ dumbbell we obtain: $m \approx 226m_{0}$, $I_{\perp} \approx 3.59\cdot10^{3} m_{0}\sigma^{2}$, and $I_{\parallel} \approx 814m_{0}\sigma^{2}$.

\subsection{\label{sub:LB_para}Lattice-Boltzmann Parameters}

The raspberry particles were coupled to an LB fluid using the coupling described in Section~\ref{sub:LBmeth}. We did \underline{not} employ the coupling scheme of Refs.~\cite{Mackay13a,Mackay13b}, since our method turned out to work sufficient well for the long-time properties without modifications to the Ahlrichs and D{\"u}nweg LB coupling. We used a graphics processing unit (GPU) based LB solver,~\cite{roehm12} which is attached to the MD software \textit{ESPResSo}.~\cite{limbach06a,arnold13a} The GPU variant of LB implemented in \textit{ESPResSo} utilizes a D3Q19 lattice and a fluctuating multi-relaxation time (MRT) collision operator.~\cite{dhumieres02} This fluctuating LB model was introduced first by Adhikari~\textit{et al.}~\cite{adhikari05} and later validated by D{\"u}nweg~\textit{et al.}~\cite{schiller07,schiller09} 

To keep our result as general as possible, we set the density of the fluid to $\rho = 1 m_{0}\sigma^{-3}$, the lattice spacing to $1\sigma$, the time step to $\Delta t = 0.005 \tau$, the (kinematic) viscosity to $\nu = 1 \sigma^{2}\tau^{-1}$, the bare particle-fluid friction to $\zeta_{0} = 25 m_{0}\tau^{-1}$, and the strength of the fluctuations to $k_{\mathrm{B}}T = 0.01\epsilon$, unless otherwise specified. Here, we chose neither to optimize our parameters for the most accurate reproduction of hydrodynamic interactions, nor to match a specific experimental system of interest \textit{via} telescoping.~\cite{louis06,louis10} We instead simply chose to use parameters that are in the regime, where LB reproduces hydrodynamic effects for colloids reasonably well and is sufficiently stable to use the (float-precision) GPU algorithm, as we will further discuss in Section~\ref{sub:dimnum}. The low amplitude of the fluctuations in the thermalized LB is to allow averaging over long times without noise dominating our results. This will become more clear when we discuss these results and prove the importance for the thermal averaging performed in Part II~\cite{degraaf15}.

\subsection{\label{sub:exper}Hydrodynamic Experiments}

To assess the quality of the raspberry approximation in modeling the hydrodynamic properties of a colloid we performed several experiments. We use the term `quiescent' to describe an un-thermalized (non-fluctuating, deterministic) LB fluid. Below we specify the experiments performed for raspberry particles in a simple cubic lattice, \textit{i.e.}, a cubic simulation box of length $L$ with periodic boundary conditions. In all experiments the particle was initialized in the center of the box.

\begin{figure}[!htb]
\begin{center}
\includegraphics[scale=1.0]{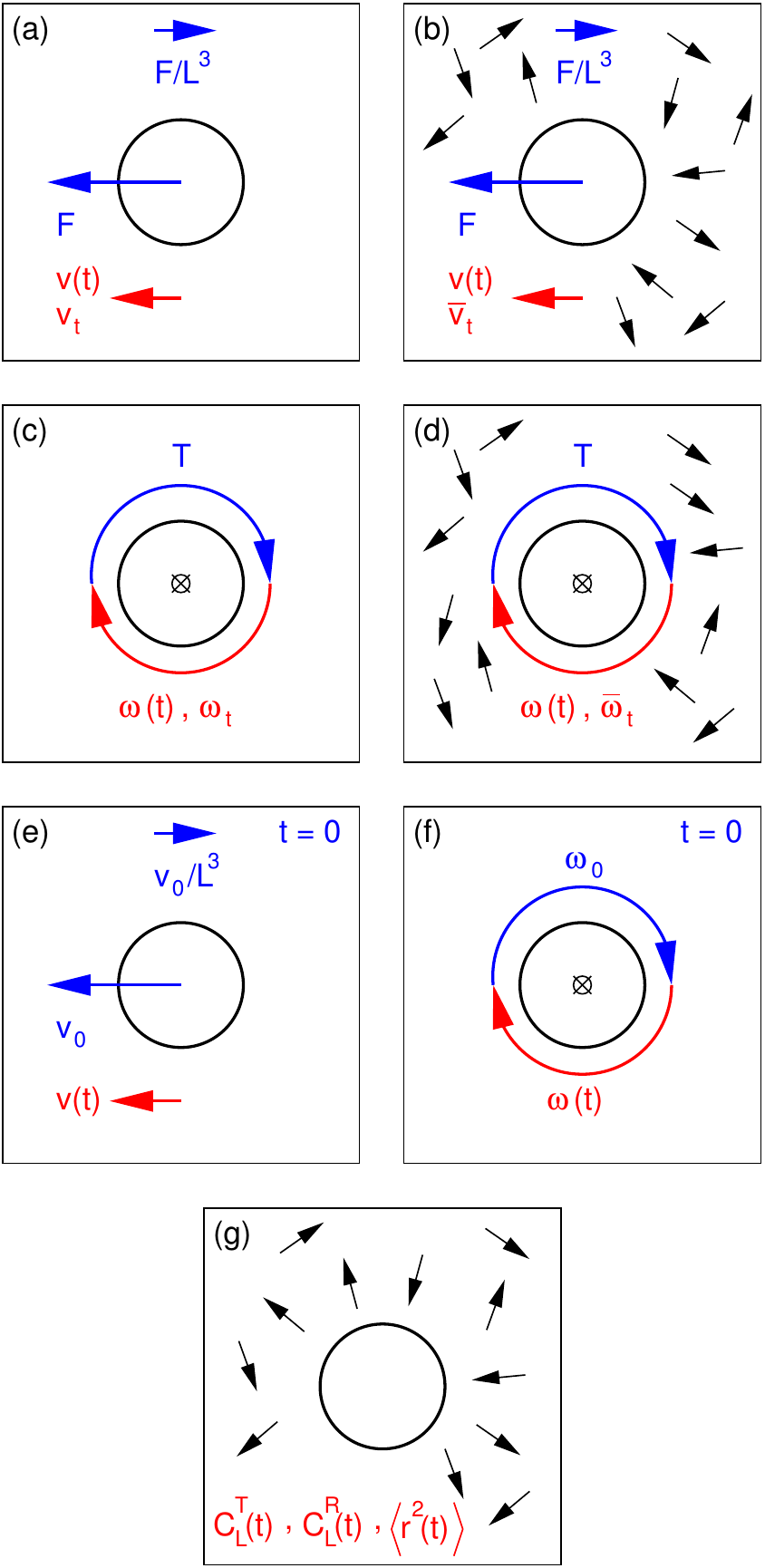}
\end{center}
\caption{\label{fig:cube}(color online) Visualization of the various hydrodynamic experiments carried out in a cubic box of length $L$ with periodic boundary conditions. A two-dimensional (2D) representation is given here. The blue arrows and symbols denote quantities applied to the fluid and raspberry, the red arrows and symbols indicate measured quantities. The black arrows indicate a thermalized fluid. We refer to the text for a description of the experiments, as well as the applied and measured quantities.}
\end{figure}

\begin{itemize}

\item A \textit{force} experiment in a \underline{quiescent} fluid, see Fig.~\ref{fig:cube}(a). A constant force $\mathbf{F}$ was applied to the particle (typically along one of the box axes) and a counter force of $-\mathbf{F}/L^{3}$ was applied homogeneously to the fluid to ensure that there is no motion of the center of mass, \textit{i.e.}, no net transfer of momentum to the system. Not applying this counter force would result in an acceleration of the colloid \textit{via} the fluid flow that builds up, as momentum is continuously pumped into the system. The resulting time-dependent velocity $\mathbf{v}(t)$ and steady-state (terminal) velocity $\mathbf{v}_{t}$ were measured and used to determine the translational mobility 
\begin{equation}
\label{eq:ES} \mu^{T}_{L} = \frac{\vert \mathbf{v}_{t} \vert}{\vert \mathbf{F} \vert} .
\end{equation} 
To establish the steady-state velocity it proved necessary to average over several (very small) oscillations in the velocity $\mathbf{v}(t)$ that are caused by lattice-discretization artifacts.

\item A \textit{force} experiment in a \underline{thermalized} fluid, see Fig.~\ref{fig:cube}(b). The set-up is the same as for the experiment in Fig.~\ref{fig:cube}(a). However, the system was first equilibrated until a steady-state emerged and the particle fluctuated with the proper thermal velocity distribution. During the production run, $\mathbf{v}(t)$ was averaged to determine the average steady-state velocity $\bar{\mathbf{v}}_{t}= \langle \mathbf{v}(t) \rangle$, where $\langle \cdot \rangle$ denotes the time average. This allowed us to determine the time-averaged translational mobility
\begin{equation}
\label{eq:ESav} \bar{\mu}^{T}_{L} = \frac{\vert \bar{\mathbf{v}}_{t} \vert}{\vert \mathbf{F} \vert} .
\end{equation} 

\item A \textit{torque} experiment in a \underline{quiescent} fluid, see Fig.~\ref{fig:cube}(c). A constant torque $\mathbf{T}$ was applied to the particle (typically along one of the box axes). The resulting time-dependent angular velocity $\boldsymbol{\omega}(t)$ and steady-state angular velocity $\boldsymbol{\omega}_{t}$ were measured and used to determine the rotational mobility
\begin{equation}
\label{eq:ESrot} \mu^{R}_{L} = \frac{\vert \boldsymbol{\omega}_{t} \vert}{\vert \mathbf{T} \vert}.
\end{equation} 
There is no need to apply a `back torque density' to the fluid in this experiment, as the periodic boundary conditions do not allow the fluid to develop a net rotation. Here, averaging of the oscillations in $\boldsymbol{\omega}(t)$ due to lattice artifacts also proved necessary.

\item A \textit{torque} experiment in a \underline{thermalized} fluid, see Fig.~\ref{fig:cube}(d). The set-up is the same as for the experiment in Fig.~\ref{fig:cube}(c). However, the system was first equilibrated until a steady-state emerged and the particle fluctuated with the proper thermal distribution. During the production run, $\boldsymbol{\omega}(t)$ was averaged to determine the average steady-state angular velocity $\bar{\boldsymbol{\omega}}_{t} = \langle \boldsymbol{\omega}(t) \rangle$. This allowed us to determine the time-averaged rotational mobility
\begin{equation}
\label{eq:ESavrot} \bar{\mu}^{R}_{L} = \frac{\vert \bar{\boldsymbol{\omega}}_{t} \vert}{\vert \mathbf{T} \vert} .
\end{equation} 

\item A \textit{velocity} experiment in a \underline{quiescent} fluid, see Fig.~\ref{fig:cube}(e). An instantaneous velocity $\mathbf{v}_{0}$ was imparted onto the particle at $t = 0$ and an instantaneous counter velocity of $-\mathbf{v}_{0}/L^{3}$ was applied homogeneously (at the same time) to the fluid to ensure zero net motion of the system. The resulting time-dependent velocity $\mathbf{v}(t)$ was measured. This quantity can be related to a non-dimensionalized velocity auto-correlation function (VACF)
\begin{eqnarray}
\label{eq:vacf} C^{T}_{L}(t) & = & \frac{ \mathbf{v}_{0} \cdot \mathbf{v}(t) }{\vert \mathbf{v}_{0} \vert^{2}} .
\end{eqnarray}

\item An \textit{angular velocity} experiment in a \underline{quiescent} fluid, see Fig.~\ref{fig:cube}(f). An instantaneous angular velocity $\boldsymbol{\omega}_{0}$ was imparted onto the particle at $t = 0$. The resulting time-dependent angular velocity $\boldsymbol{\omega}(t)$ was measured. This quantity can be related to a non-dimensionalized angular velocity auto-correlation function (AVACF)
\begin{eqnarray}
\label{eq:avacf} C^{R}_{L}(t) & = & \frac{\boldsymbol{\omega}_{0} \cdot \boldsymbol{\omega}(t)}{ \vert \boldsymbol{\omega}_{0} \vert^{2}} .
\end{eqnarray}

\item An \textit{auto-correlation} experiment in a \underline{thermalized} fluid, see Fig.~\ref{fig:cube}(g). The system was equilibrated until the particle fluctuated with the proper thermal distribution. The (A)VACF and the mean square displacement (MSD) were measured using the multiple-tau correlator in \textit{ESPResSo}.~\cite{ramirez10} For the (A)VACF the (angular) velocity in the co-rotating frame was averaged. The $C^{T}(t) \equiv \langle \mathbf{v}(t) \cdot \mathbf{v}(t+\tau) \rangle$ and $C^{R}(t) \equiv \langle \boldsymbol{\omega}(t) \cdot \boldsymbol{\omega}(t+\tau) \rangle$ that follow from the thermal experiments differ slightly from those in Eqs.~\cref{eq:vacf}~and~\cref{eq:avacf}, because $C^{T}(0) = 3 k_{\mathrm{B}} T / m$ and $C^{R}(0) = 3 k_{\mathrm{B}} T / I$, as a consequence of the equipartition theorem. This allows us to compute the translational and rotational mobility, respectively, \textit{via} the Green-Kubo relation
\begin{equation}
\label{eq:GK} \mu^{X}_{L} = \frac{1}{3 k_{\mathrm{B}} T} \int_{0}^{\infty} C^{X}(t) \ud t,
\end{equation}
where the factor $1/3$ is used for spherical particles only and $X$ can be either $T$ or $R$.~\cite{Hauge73} The relations for anisotropic particles are similar, but slightly more involved, since the dot product for the (A)VACF is replaced by the dyadic product.

\end{itemize}

\subsection{\label{sub:dimnum}Dimensionless Numbers for the Fluid Properties}

In the above experiments, care was taken to ensure that the particle remained in the low translational Reynolds number regime
\begin{equation}
\label{eq:Retrans} Re^{T} = \frac{v R}{\nu} \ll 1,
\end{equation}
with $v$ the maximum/typical velocity. This implies that we can compare it to analytic and numerical results obtained by solving the Stokes equations, as will be discussed further in Section~\ref{sec:result}. For the colloid radius $R = 3.0\sigma$ and our value of the kinematic viscosity, we ensured that the maximum particle velocity remained under $0.15\sigma\tau^{-1}$, for which $Re^{T} < 0.5$. However, this value was only attained in the \textit{velocity} and \textit{auto-correlation} experiments for the first time step. For $t \gg 1 \tau$ and in the other experiments, the Reynolds number remained smaller than 0.1. Similarly, the rotational Reynolds number
\begin{equation}
\label{eq:Rerot} Re^{R} = \frac{\omega R^{2}}{\nu},
\end{equation}
with $\omega$ the maximum angular velocity, remained small: $Re^{R} < 0.7$, but typically smaller than $0.1$. For the other radii that we considered, the maximum value of the Reynolds number was kept smaller. 

There are a number of relevant parameters to describe the hydrodynamic properties of our system. For the thermalized LB fluid, we can define the P{\'e}clet and Schmidt number of the particle, and the Boltzmann number of the fluid. In both quiescent and thermalized LB fluids, we can determine the Mach number. Finally, the coupling of the raspberry particles to the fluid can be described by the Immersion number and the Screening ratio. We will determine these numbers next.

The translational and rotational P{\'e}clet numbers are defined as 
\begin{eqnarray}
\label{eq:PeT} Pe^{T} & = & \frac{v R}{D^{T}_{0}} ; \\
\label{eq:PeR} Pe^{R} & = & \frac{\omega}{D^{R}_{0}}.
\end{eqnarray}
For the thermalized force and torque experiments, see Figs.~\ref{fig:cube}(b,d), we obtain values of $Pe^{T} \approx 1.5\cdot10^{3}$ and $Pe^{R} \approx 5.0\cdot10^{2}$ for the $R = 3.0\sigma$ raspberry. We did not carry out similar thermalized experiments for $R \ne 3.0\sigma$. If we use the thermal velocity for the auto-correlation experiment, see Fig.~\ref{fig:cube}(g), then we arrive at $Pe^{T} \approx 2.5\cdot10^{2}$ and $Pe^{R} \approx 5.0\cdot10^{2}$. The large value of the P{\'e}clet number indicates that our results are in a regime that is dominated by advective flow, rather than by diffusion. That is, our thermalized results can be readily compared to those of the quiescent (deterministic) experiments.

The Schmidt number of the particles measures the relative importance of diffusive and advective transport and is defined as 
\begin{equation}
\label{eq:Sc} Sc = \frac{\nu}{D^{T}_{0}},
\end{equation}
where $\nu$ is the kinematic viscosity, as before. We obtain $Sc = 5.6\cdot10^{3}$ for the $R = 3.0\sigma$ colloid and a thermal fluctuation strength of $k_{\mathrm{B}}T = 0.01 \epsilon$. This value of the Schmidt number is quite high, compared to the typical value of $Sc \approx 10$ in LB simulations. However, it is necessary to use such high numbers to access the regime in which the momentum diffusion in the fluid dominates the diffusive transport of the particles. This allows for the accurate measurement of hydrodynamic interactions in confining geometries, see Ref.~\cite{degraaf15}.

The Boltzmann number $Bo$ of the LB fluid, which indicates the level of coarse graining, is defined as
\begin{equation}
\label{eq:Bo} Bo = \frac{\left( \langle v_{i}^{2} \rangle - \langle v_{i} \rangle^{2} \right)^{1/2} }{ \langle v_{i} \rangle },
\end{equation}
where $v_{i}$ is the speed of the LB fluid at a given node in the fluid.~\cite{dunweg09} By averaging over $10^{3}$ LB nodes for the parameters that we used, we obtain $Bo = 0.82$. For $Bo = 1$ (the maximum value) the model is fully microscopic, whereas for $Bo = 0$ the model is entirely deterministic. For this value of the Boltzmann number we are in an intermediate regime, with a limited level of coarse graining.

The Mach number of the LB fluid is the ratio of the particle velocity to the speed of sound and is given by 
\begin{equation}
\label{eq:Ma} Ma = \frac{v}{v_{s}},
\end{equation}
where $v_{s}$ is the speed of sound in LB 
\begin{equation}
\label{eq:LBvs} v_{s} = \sqrt{\frac{1}{3}}\left( \frac{a}{\Delta t} \right),
\end{equation}
with $a$ the lattice spacing and $\Delta t$ the time step. The prefactor depends on the shape and dimensionality of the grid (the prefactor for a D2Q9 grid is the same incidentally). For our parameters we obtain $Ma \approx 9.0\cdot10^{-4}$ for the thermalized force experiment of Fig.~\ref{fig:cube}(b) and $Ma \approx 1.5\cdot10^{-4}$ if we take $v$ to be the thermal velocity of a $R = 3.0\sigma$ colloid at $k_{\mathrm{B}}T = 0.01 \epsilon$. For all radii we obtain $Ma < 10^{-3}$.

\begin{table}
\begin{ruledtabular}
\begin{tabular}{c|c|c|c|c|c|c}
$R$          & $N_{\mathrm{tot}}$ & $\zeta_{0}$          & $\kappa_{\mathrm{X}}$ & $R^{T}_{\mathrm{P}}$ & $R^{R}_{\mathrm{P}}$ & $R^{T}_{\mathrm{P}}/R^{R}_{\mathrm{P}}$ \\
\hline
\multicolumn{7}{c}{}\\[-0.8em]
\multicolumn{7}{c}{Filled} \\
\hline
\\[-0.8em]
$5.0 \sigma$ & 1917               & $25 m_{0} \tau^{-1}$ & 32.3                  & $4.84\sigma$         & $4.85\sigma$         & 0.9986                                  \\
$4.0 \sigma$ & 1409               & $25 m_{0} \tau^{-1}$ & 30.9                  & $3.86\sigma$         & $3.87\sigma$         & 0.9985                                  \\
$3.0 \sigma$ & 925                & $35 m_{0} \tau^{-1}$ & 31.0                  & $2.90\sigma$         & $2.90\sigma$         & 0.9985                                  \\
$3.0 \sigma$ & 925                & $25 m_{0} \tau^{-1}$ & 28.9                  & $2.89\sigma$         & $2.90\sigma$         & 0.9983                                  \\
$3.0 \sigma$ & 925                & $10 m_{0} \tau^{-1}$ & 22.3                  & $2.86\sigma$         & $2.87\sigma$         & 0.9971                                  \\
$3.0 \sigma$ & 603                & $25 m_{0} \tau^{-1}$ & 23.4                  & $2.86\sigma$         & $2.87\sigma$         & 0.9973                                  \\
$3.0 \sigma$ & 403                & $25 m_{0} \tau^{-1}$ & 19.1                  & $2.83\sigma$         & $2.84\sigma$         & 0.9960                                  \\
$3.0 \sigma$ & 303                & $25 m_{0} \tau^{-1}$ & 16.6                  & $2.80\sigma$         & $2.82\sigma$         & 0.9948                                  \\
$3.0 \sigma$ & 253                & $25 m_{0} \tau^{-1}$ & 15.1                  & $2.78\sigma$         & $2.80\sigma$         & 0.9938                                  \\
$2.5 \sigma$ & 248                & $25 m_{0} \tau^{-1}$ & 16.4                  & $2.34\sigma$         & $2.35\sigma$         & 0.9947                                  \\
$2.0 \sigma$ & 245                & $25 m_{0} \tau^{-1}$ & 18.2                  & $1.88\sigma$         & $1.89\sigma$         & 0.9957                                  \\
\hline
\multicolumn{7}{c}{}\\[-0.8em]
\multicolumn{7}{c}{Hollow} \\
\hline
\\[-0.8em]
$5.0 \sigma$ & 594                & $25 m_{0} \tau^{-1}$ & 18.0                  & $4.93\sigma$         & $4.95\sigma$         & 0.9953                                  \\
$4.0 \sigma$ & 442                & $25 m_{0} \tau^{-1}$ & 17.3                  & $3.94\sigma$         & $3.96\sigma$         & 0.9950                                  \\
$3.0 \sigma$ & 203                & $35 m_{0} \tau^{-1}$ & 14.5                  & $2.94\sigma$         & $2.96\sigma$         & 0.9928                                  \\
$3.0 \sigma$ & 925                & $25 m_{0} \tau^{-1}$ & 28.9                  & $2.98\sigma$         & $2.99\sigma$         & 0.9982                                  \\
$3.0 \sigma$ & 203                & $25 m_{0} \tau^{-1}$ & 13.5                  & $2.93\sigma$         & $2.95\sigma$         & 0.9918                                  \\
$3.0 \sigma$ & 203                & $10 m_{0} \tau^{-1}$ & 10.4                  & $2.88\sigma$         & $2.92\sigma$         & 0.9861                                  \\
$2.5 \sigma$ & 140                & $25 m_{0} \tau^{-1}$ & 12.3                  & $2.42\sigma$         & $2.45\sigma$         & 0.9900                                  \\
$2.0 \sigma$ & 90                 & $25 m_{0} \tau^{-1}$ & 11.0                  & $1.93\sigma$         & $1.95\sigma$         & 0.9876                                  \\
\end{tabular}
\end{ruledtabular}
\caption{\label{tab:kappa}The screening-ratio related properties for the various raspberry particles studied in this manuscript. From left to right, the columns give: the imposed radius $R$, the total number of coupling points in the raspberry $N_{\mathrm{tot}}$, the value of the bare friction coefficient $\zeta_{0}$, the screening ratio $\kappa_{\mathrm{X}}$ (`X' is either `F' or `H'), the translational hydrodynamic radius for the porous sphere $R^{T}_{\mathrm{P}}$,~\cite{Debye48,Felderhof75b} the rotational hydrodynamic radius for the porous sphere $R^{R}_{\mathrm{P}}$,~\cite{Felderhof75a} and finally the ratio of these radii.}
\end{table}

The immersion number, which measures the relative density of the MD beads, is defined as 
\begin{equation}
\label{eq:im} \displaystyle In = \left( 1 + \frac{m_{0}}{\rho \sigma^{3}} \right)^{-1},
\end{equation}
where $m_{0}$ is the mass of a single MD bead. It should be noted that the individual (virtual) MD beads, which make up the raspberry, have unit mass ($m_{\mathrm{bead}} = 1.0m_{0}$). Only the central bead, to which the other beads couple and which holds the properties of the entire colloid, has a different mass on the MD level. For our choice of parameters we obtain $In = 0.5$, which corresponds to a neutrally buoyant object. 

The screening ratio~\cite{Debye48,Felderhof75a,Felderhof75b} for a filled sphere is a measure for the porosity of an object and it is given by
\begin{equation}
\label{eq:srfill} \kappa_{\mathrm{F}} = R \sqrt{\displaystyle \frac{\rho_{\mathrm{F}} \zeta_{\mathrm{eff}}}{\eta}} ,
\end{equation}
where $\eta$ is the dynamic viscosity, and
\begin{equation}
\label{eq:rhoF} \rho_{\mathrm{F}} = \frac{ 3 N_{\mathrm{tot}}}{4\pi R^{3}},
\end{equation}
is the density of coupling beads in the sphere (assuming a uniform distribution -- which is an acceptable approximation for our fillings), $\zeta_{\mathrm{eff}}$ is the effective single particle fluid-coupling~\cite{dunweg09}
\begin{equation}
\label{eq:feff} \zeta_{\mathrm{eff}} = \left( \frac{1}{\zeta_{0}} + \frac{0.04}{\eta a} \right)^{-1},
\end{equation}
with $\zeta_{0}$ the imposed LB fluid friction, $0.04$ a prefactor for the D3Q19 grid, the value of which we measured, and $a$ is the LB grid spacing. For the various particles that we used the screening ratio is given in Table~\ref{tab:kappa}. For a hollow sphere~\cite{Debye48,Felderhof75b} the screening ratio is defined by
\begin{equation}
\label{eq:srholl} \kappa_{\mathrm{H}} = \sqrt{ 3R \frac{\rho_{\mathrm{H}} \zeta_{\mathrm{eff}}}{\eta}} ,
\end{equation}
where
\begin{equation}
\label{eq:rhoH} \rho_{\mathrm{H}} = \frac{N_{\mathrm{tot}}}{4\pi R^{2}},
\end{equation}
is the density of beads on the surface of the sphere assuming a uniform distribution. 

The screening radio gives insight into the match between the translational and rotational hydrodynamic radius of the particle, as predicted by porous particle theory.~\cite{Debye48,Felderhof75a,Felderhof75b} For the filled sphere the following hydrodynamic radii are expected
\begin{eqnarray}
\label{eq:FporRT} R^{T}_{\mathrm{P}} & = & R \frac{\left(1 - \displaystyle \frac{\tanh(\kappa_{\mathrm{F}})}{\kappa_{\mathrm{F}}} \right) }{1 + \displaystyle \left( \frac{3}{2 \kappa_{\mathrm{F}}^{2}} \right) \left( 1 - \displaystyle \frac{\tanh(\kappa_{\mathrm{F}})}{\kappa_{\mathrm{F}}} \right)} ; \\
\label{eq:FporRR} R^{R}_{\mathrm{P}} & = & R \left( 1 - 3 \frac{\coth(\kappa_{\mathrm{F}})}{\kappa_{\mathrm{F}}} + \frac{3}{\kappa_{\mathrm{F}}^{2} } \right)^{1/3},
\end{eqnarray}
where $R^{T}_{\mathrm{P}}$~\cite{Felderhof75b} is the translational hydrodynamic radius and $R^{R}_{\mathrm{P}}$~\cite{Felderhof75a} is the rotational one. For the hollow shell the expressions are~\cite{Debye48,Felderhof75a,Felderhof75b}
\begin{eqnarray}
\label{eq:HporRT} R^{T}_{\mathrm{P}} & = & R \frac{\kappa_{\mathrm{H}}^{2}}{\displaystyle \frac{9}{2} + \kappa_{\mathrm{H}}^{2}} ; \\
\label{eq:HporRR} R^{R}_{\mathrm{P}} & = & R \left( \frac{\kappa_{\mathrm{H}}^{2}}{9 + \kappa_{\mathrm{H}}^{2}} \right)^{1/3}.
\end{eqnarray}
The values of these radii are given in Table~\ref{tab:kappa}. It should be noted that the ratio of the two radii (translational and rotational) is only equal when $\kappa_{\mathrm{X}} \uparrow \infty$ (`X' is either `F' or `H'). For $\kappa_{\mathrm{X}} \downarrow 0$ the translational hydrodynamic radius is much smaller than the rotational one. We have added the values of these ratios to Table~\ref{tab:kappa}. In our analysis and discussion, see Sections~\ref{subb:porous} and~\ref{sec:disc}, we relate the insights of porous-particle theory to our results for the raspberry particle.

\subsection{\label{sub:notation}Notations Used throughout this Manuscript}

In this section, we summarize the notations used in this manuscript. This will aid the reader in going through the text, as many of the notations are necessarily similar.

\begin{itemize}
\setlength{\itemsep}{0pt}

\item $L$, the box length of the cubic box with periodic boundary conditions.

\item $R_{h}^{T}$, the effective hydrodynamic radius obtained by extrapolating translational mobility measurements, see Figs.~\ref{fig:cube}(a,b,e,g), for the limit of box length $L\uparrow\infty$. The subscript $h$ is used to differentiate $R_{h}$ from the bead-to-center distance of the raspberry's coupling points $R$.

\item $R_{h}^{R}$, the effective hydrodynamic radius obtained by extrapolating rotational mobility measurements, see Figs.~\ref{fig:cube}(c,f,g), for the limit of box length $L\uparrow\infty$.

\item $C^{T}_{L}(t)$, the velocity auto-correlation function (VACF) for translational movement in a cubic box of length $L$ with periodic boundary conditions, see Figs.~\ref{fig:cube}(e,g) and Eq.~\cref{eq:vacf}.

\item $C^{R}_{L}(t)$, the angular velocity auto-correlation function (AVACF) for rotation in a cubic box of length $L$ with periodic boundary conditions, see Figs.~\ref{fig:cube}(f,g) and Eq.~\cref{eq:avacf}.

\item $\mu^{T}_{L}(t)$, the time-dependent translational mobility in a cubic box of length $L$ with periodic boundary conditions, see Fig.~\ref{fig:cube}(a). When the time dependence is dropped, the limit $t \uparrow \infty$ has been taken.

\item $\mu^{R}_{L}(t)$, the time-dependent rotational mobility in a cubic box of length $L$ with periodic boundary conditions, see Fig.~\ref{fig:cube}(c). When the time dependence is dropped, the limit $t \uparrow \infty$ has been taken.

\item $\bar{\mu}^{T}_{L}$, the time-averaged translational mobility resulting from the thermal force experiment, see Fig.~\ref{fig:cube}(b).

\item $\bar{\mu}^{R}_{L}$, the time-averaged rotational mobility resulting from the thermal torque experiment, see Fig.~\ref{fig:cube}(d).

\item $\mu^{T}_{0}$, the bulk translational mobility, which follows from the limit $L \uparrow \infty$ of $\mu^{T}_{L}$.

\item $\mu^{R}_{0}$, the bulk rotational mobility, which follows from the limit $L \uparrow \infty$ of $\mu^{R}_{L}$.

\item $f$, the fractional deviation between two results.

\end{itemize}

\section{\label{sec:result}Results}

In this section, we discuss the results that we obtained by performing the simulations and numerical calculations outlined in Section~\ref{sec:methods}. We have split this section into two parts: one for the sphere and one for the dumbbell. These parts are further subdivided according to the nature of the experiments.

\subsection{\label{sub:SCsphere}Sphere in a Simple Cubic Crystal}

\subsubsection{\label{subb:VACFsphere}The (Angular) Velocity Auto-Correlation Function}

\begin{figure*}
\begin{center}
\includegraphics[scale=1.0]{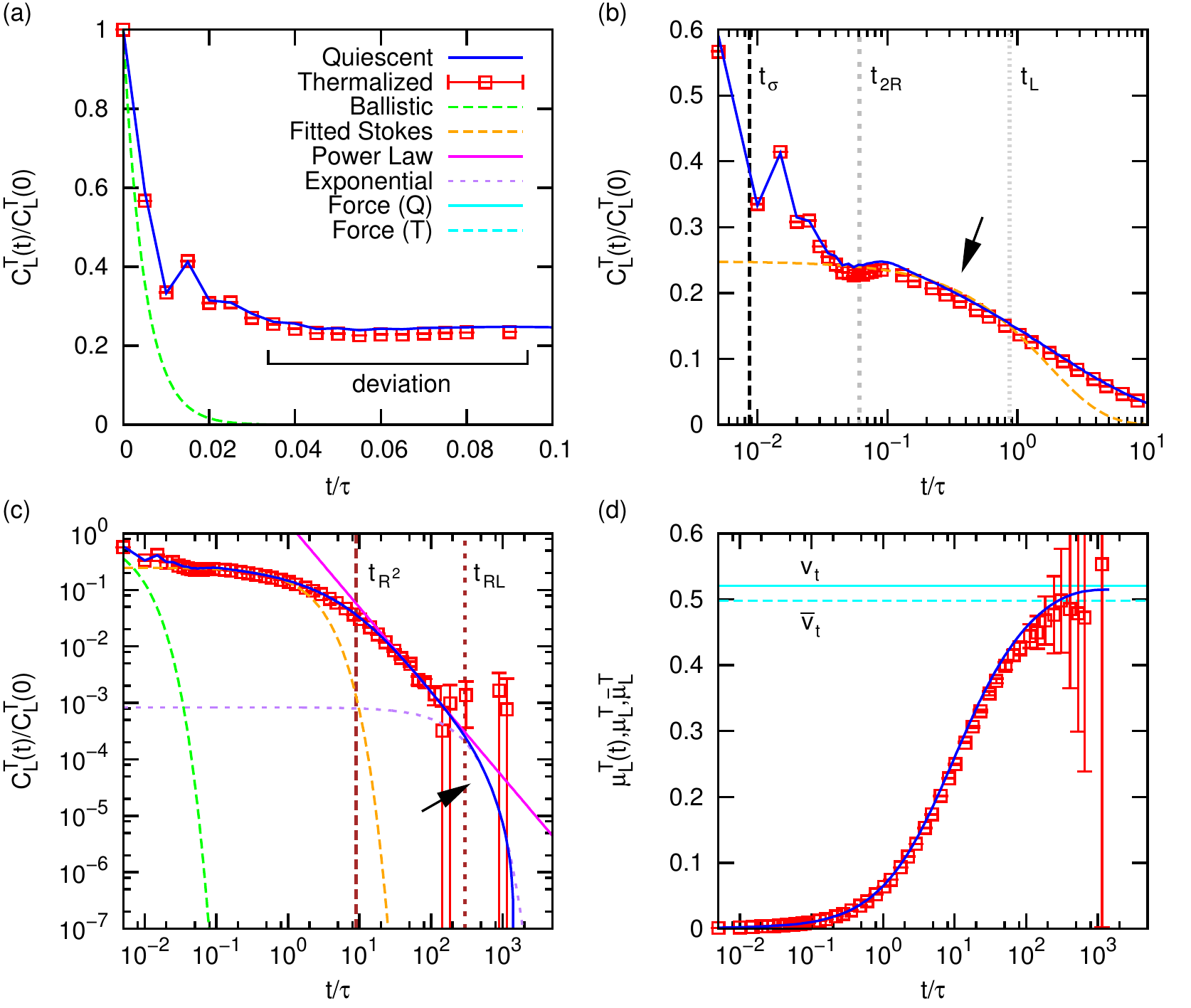}
\end{center}
\caption{\label{fig:vacf_filled}(color online) The velocity auto-correlation function (VACF) $C_{L}^{T}(t)$ as a function of the time $t$ expressed in the MD time unit $\tau$. The graphs show results for a filled raspberry of radius $R = 3.0\sigma$ in a box of length $L = 100.0\sigma$, with LB parameters as given in the text. (a) The initial decay of the VACF. The red squares with error bars show the result for a thermalized LB, the blue solid curve gives the result of a quiescent experiment, and the green dashed line shows the predicted unphysical-coupling decay inherent to LB. (b) Log-linear plot of the initial and intermediate decay of the VACF. The dashed orange curve gives the expected Stokes' decay. An arrow indicates the position where the correspondence is reasonable. The vertical lines indicate the time for sound waves to propagate through the system over certain lengths: one lattice spacing ($t_{\sigma}$, dashed black), roughly the inter-bead separation; the raspberry's hydrodynamic diameter ($t_{2R}$, dashed gray); and the box length ($t_{L}$, dotted gray). (c) Log-log plot of the long-time decay. The magenta line shows the power-law decay. The unphysical coupling, fitted Stokes', and final exponential-decay (purple dashed, indicated by an arrow) curves are shown for completeness. The two brown vertical lines indicate the time needed for viscous dissipation over certain lengths: the radius of the sphere ($t_{R^{2}}$, dashed), and the product of the sphere radius $R$ with the box length $L$ ($t_{RL}$, short dashes). (d) The time-dependent Green-Kubo value of the translational mobility $\mu^{T}_{L}(t)$ obtained from the quiescent (blue solid curve) and thermalized (red squares with error bars) LB result. The solid cyan line shows the result of a quiescent force experiment ($\mu^{T}_{L}$, derived from the terminal velocity $v_{t}$), while the dashed cyan line shows the result of a thermal-averaged force experiment ($\bar{\mu}^{T}_{L}$, from the time-averaged terminal velocity $\bar{v}_{t}$).}
\end{figure*}

Using the (quiescent) \textit{velocity} and (thermalized) \textit{auto-correlation} experiments discussed in Section~\ref{sub:exper}, see Figs.~\ref{fig:cube}(e,g), we established the VACF for a filled raspberry sphere in a cubic box of length $L=100.0\sigma$. The results are shown in Fig.~\ref{fig:vacf_filled}. From Figs.~\ref{fig:vacf_filled}(a,b,c) we observe several decay regimes that are typical for the LB simulations of the raspberry particle. In the following they will be described in more detail.

$~$
\paragraph*{Decay Regimes}
$~$\newline

\textbf{(I)} At short times there is an unphysical-coupling regime, see Fig.~\ref{fig:vacf_filled}(a), in which the VACF decays exponentially according to 
\begin{equation}
\label{eq:VACFbal} \frac{C^{T}_{L}(t)}{C^{T}_{L}(0)} = \exp \left( -\frac{ N_{\mathrm{tot}} \zeta_{0} }{ m } t \right), 
\end{equation} 
with $N_{\mathrm{tot}}$ the total number of beads, $\zeta_{0}$ the bare friction coefficient, and $m$ the particle's (bare) mass.~\cite{lobaskin04} The existence of this regime can be attributed to the fluid not co-moving with the velocity of the beads (raspberry particle). That is, the MD beads interact with the stationary fluid only through a regular Langevin-type friction -- the velocity of the fluid is essentially zero during these time steps.

The expected (unphysical) decay of Eq.~\cref{eq:VACFbal} is indicated in Fig.~\ref{fig:vacf_filled}(a) and matches reasonably well with the observed initial decay. However, the result deviates even in the first and second time step, signifying the onset of proper coupling. This is in agreement with the recent observations in the MPCD simulations of Ref.~\cite{poblete14}, where this deviation from the expected unphysical decay was also attributed to the onset of hydrodynamic correlations. Finally, note that there is a small deviation between the thermalized LB result and the quiescent VACF when $t > 0.03\tau$, to which we will come back later.

From the above it is thus clear that the no-slip boundary condition at the surface of the raspberry is violated at short times, even taking the finite compressibility of the LB fluid into account. Moreover, the expected decay for a porous colloid~\cite{Felderhof14} is \underline{not} captured by the raspberry with the Ahlrichs and D{\"u}nweg coupling.~\cite{ahlrichs99} This is a problem inherent to the LB method.~\cite{lobaskin04,chatterji05} The modified coupling scheme by Mackay~\textit{et al.}~\cite{Mackay13a} purportedly remedies this problem, we will come back to this in Section~\ref{sec:disc}.

\textbf{(II)} \uuline{The Decay.} At intermediate times there is a regime, in which the VACF decays exponentially according to Stokes' prediction
\begin{equation}
\label{eq:VACFSto}  \frac{C^{T}_{L}(t)}{C^{T}_{L}(0)} \propto \exp \left( -\frac{ 6 \pi \eta R }{ m } t \right) .
\end{equation}
Here, we used the proportionality symbol, since the unphysical initial decay makes it impossible to establish an analytic prefactor for the onset of this regime in fluid-particle coupling. The regime appears because the hydrodynamic coupling between the raspberry particle and the surrounding fluid is now fully established.~\cite{lobaskin04} It should be noted that in Ref.~\cite{lobaskin04} the mass in the denominator was taken to be $m^{\ast}$, where $m^{\ast}$ is the `virtual' mass.~\cite{Zwanzig75} This virtual mass is the particle mass $m$ plus half of the displaced fluid mass; $m^{\ast} = 3m/2$ in our case. We will come back to this shortly. 

The applicability of Stokes' prediction for our numerical results can be seen in Fig.~\ref{fig:vacf_filled}(b), where a Stokes-type decay has been fitted to our data. The agreement is not very convincing. The curve does not match the Stokes' trend well. However, the agreement between the bare-mass prediction of Eq.~\cref{eq:VACFSto} is superior to the one in which the virtual mass is used (not shown here). The latter type of decay was originally suggested by Lobaskin and D{\"u}nweg.~\cite{lobaskin04} The superiority of the bare-mass result could be reasonable since Felderhof~\cite{Felderhof14} has shown that for a porous sphere the $m^{\ast}$-related decay regime is absent in the high-frequency limit. Unfortunately, it is unclear whether our simulations are sufficiently close to this limit. In addition, in the limit where the viscous coupling constant goes to infinity before the frequency, the virtual mass decay is present.~\cite{Felderhof14} The fact that the high-frequency porous sphere solution of Felderhof~\cite{Felderhof14} does not match better in the Stokes-type regime, makes for a slightly academic discussion, since such comparison is hindered by the presence of the unphysical decay. 

\uuline{Characteristic Times.} We have indicated three characteristic times related to sound propagation in the LB in Fig.~\ref{fig:vacf_filled}(b). The three times are $t_{\sigma} = \sigma/v_{s}$, $t_{2R} = 2R/v_{s}$, and $t_{L} = L/v_{s}$, \textit{i.e.}, the time required for sound waves to propagate one lattice spacing, the diameter of the raspberry, and the length of the box, respectively. Here, $v_{s}$ is the speed of sound, as defined in Eq.~\cref{eq:LBvs}. We will now discuss the relevance of these times.

For the filled sphere, in which the MD beads are roughly $1.0\sigma$ apart, we find possible signatures of the propagation of sound between the MD beads, as can be inferred from the short-time oscillations. The first dip in the VACF roughly coincides with $t_{\sigma} \approx 8.7 \cdot 10^{-3}\tau$, as indicated by the black dashed line in Fig.~\ref{fig:vacf_filled}(b). These oscillations may also be related to the magnitude of the effective friction that the added coupling points in the interior bring about. At the time it takes sound to propagate the diameter of the sphere ($t_{2R} \approx 6.1 \cdot 10^{-2} \tau$), we find a small dip in the VACF, see the dashed gray line in Fig.~\ref{fig:vacf_filled}(b). This dip is similar to the one observed in Ref.~\cite{poblete14} and is caused by the compressibility of the LB fluid.~\cite{Zwanzig75}

Note that the Stokesian regime of decay appears to be delimited by the time it takes sound to travel the distance of the box ($t_{L} \approx 0.87 \tau$, dotted gray line in Fig.~\ref{fig:vacf_filled}(b)). However, for our specific choice of parameters, this time is close to the viscous time it takes momentum to diffuse by one colloidal radius $t_{R^{2}} = \rho R^{2} / \eta = 9.0 \tau$. This \underline{viscous time} is the relevant time scale for the development of hydrodynamic memory effects.~\cite{Zwanzig75,Hauge73} We have a stricter separation of sonic and viscous time scales than in Refs.~\cite{belushkin11,poblete14}, \textit{i.e.}, $t_{R^{2}}/t_{2R} \gg 1$. Therefore, our results do not display sound undulations (back tracking) in the long-time power-law regime.

\textbf{(III)} After a sufficiently long time, the hydrodynamic interactions with the surrounding fluid result in a persistence of the velocity (non-exponential decay) as the vorticity diffuses away from the particle. These hydrodynamic memory effects lead to an algebraic decay of the (A)VACF; the so-called `long-time tail'.~\cite{hansen86} This decay has the following form
\begin{eqnarray}
\label{eq:VACFpowtra} \frac{C^{T}_{L}(t)}{C^{T}_{L}(0)} & = & \frac{1}{12} m \sqrt{\rho} (\pi \eta t)^{-3/2} \mathcal{H}(R,L) ; \\
\label{eq:VACFpowrot} \frac{C^{R}_{L}(t)}{C^{R}_{L}(0)} & = & \pi I \sqrt{\rho} (4 \pi \eta t)^{-5/2} \mathcal{H}(R,L),
\end{eqnarray}
for the translational and rotational motion, respectively.~\cite{Hauge73,Zwanzig75,hinch75,cichocki98} \textbf{N.B. These are the 3D auto-correlation functions, which are normalized. This was unclear in our J. Chem. Phys. publication.}  Here, $\mathcal{H}(R,L)$ is the Hasimoto scaling expression~\cite{hasimoto59}
\begin{equation}
\label{eq:hasi} \mathcal{H}(R,L) = 1 - 2.837 \left( \frac{R}{L} \right)  + \frac{4\pi}{3}\left( \frac{R}{L} \right)^{3} .
\end{equation}
Figure~\ref{fig:vacf_filled}(c) shows the power-law decay for the translational motion more clearly. The correspondence with the quiescent data is excellent, we obtain a match for both the prefactor and exponent via a fitting procedure that is within 1\% of the theoretical prediction. Note that within the error bar, which gives the standard error, the decay is captured by the thermalized result. The thermal data shows correspondence within the error bar, however the error bars are substantial in this regime; it was the best that could be achieved within a reasonable time frame for our choice of parameters. Only for $L \gg 30 R$ is the power-law decay more pronounced. However, larger box sizes require even longer sampling. Our result is similar to that of Refs.~\cite{lobaskin04,chatterji05}.

In Fig.~\ref{fig:vacf_filled}(c) we have indicated two times related to viscous dissipation over certain length-scale combinations: $t_{R^{2}} = 9.0 \tau$ (as before) and $t_{RL} = \rho R L / \eta = 300.0 \tau$. These two times roughly indicate the start and end of the power-law decay. In Ref.~\cite{atzberger06} it is suggested that the exponential decay that follows the power-law decay, will set in at $t_{L^{2}} = \rho L^{2} / \eta = 1.0 \cdot 10^{4} \tau$. However, from Fig.~\ref{fig:vacf_filled}(c) it is clear that this third exponential decay, which will be discussed next, sets in far sooner than this.

\textbf{(IV)} For the quiescent data, there is a third exponential decay in the data when $t > t_{RL}$, see the purple dashed line in Fig.~\ref{fig:vacf_filled}(c). Analysis shows that this decay has a small exponent that depends on the size of the simulation box. In Ref.~\cite{atzberger06} the following form for the decay is suggested
\begin{equation}
\label{eq:VACFAtz}  \frac{C^{T}_{L}(t)}{C^{T}_{L}(0)} \propto \exp \left( -\frac{ 4 \pi^{2} \eta }{ \rho L^{2} } t \right) ,
\end{equation}
which according to Ref.~\cite{atzberger06} should set for $t > t_{L^{2}}$. The exponent comes from the smallest positive value over all potential wave numbers that fit the geometry of the box. We observe that the decay sets in more quickly in our simulation, namely around $t \approx t_{RL}$. Fitting for the value of the exponent we obtain $4.7\cdot10^{-3} \tau^{-1}$, whereas the form in Eq.~\cref{eq:VACFAtz} yields $3.9\cdot10^{-3}\tau^{-1}$. There is a difference between these two factors of about 20\%, but within the error the decay is well captured by Eq.~\cref{eq:VACFAtz} -- only our fit is shown in Fig.~\ref{fig:vacf_filled}(c). The cause of the early onset of the third exponential decay is unclear at this time.

$~$
\paragraph*{Thermal versus Quiescent}
$~$\newline

Finally, we considered the Green-Kubo relation for the VACF by integrating
\begin{equation}
\label{eq:GKad} \mu^{T}_{L}(t) = \frac{1}{3 k_{\mathrm{B}}T} \int_{0}^{t} C^{T}_{L}(t') \ud t',
\end{equation}
for the thermalized data. The expression for the quiescent data is similar. Figure~\ref{fig:vacf_filled}(d) shows the resulting time-dependent translational mobility $\mu_{L}^{T}(t)$. We obtained the value of $\mu_{L}^{T} \equiv \mu_{L}^{T}(t\uparrow\infty)=1.37\cdot10^{-2} \sigma^{2}\epsilon^{-1}\tau^{-1}$ from the quiescent data for the box of length $L=100.0\sigma$. The data for the thermalized LB has a slightly lower value than the quiescent result, which can in part be attributed to the deviation that was already present at short times.

In addition to determining $\mu_{L}^{T}$ from the VACFs we performed a quiescent and thermalized force experiment. The result is shown using the solid and dashed cyan lines in Fig.~\ref{fig:vacf_filled}(d), respectively. We arrived at a value of $\mu_{L}^{T} = 1.38\cdot10^{-2} \sigma^{2}\epsilon^{-1}\tau^{-1}$ for the quiescent data and $\bar{\mu}_{L}^{T} = 1.32\cdot10^{-2} \sigma^{2}\epsilon^{-1}\tau^{-1}$ for the thermalized data. The results from the VACF and the force experiments correspond within the error, but there is a discrepancy between the thermal and quiescent data. This deviation can be explained by the way these experiments are carried out. The counter-force applied to the fluid also acts inside of the particle, effectively modifying the total applied force, as will be further discussed in Section~\ref{subb:FSS}. 

For completeness we performed quiescent and thermal torque experiments, see Figs.~\ref{fig:cube}(c,d). From these, we obtained $\mu_{L}^{R} = 9.24\cdot10^{-3} \epsilon^{-1}\tau^{-1}$ for the quiescent data and $\bar{\mu}_{L}^{R} = 9.25\cdot10^{-3} \epsilon^{-1}\tau^{-1}$ for the thermalized data. These correspond well within the numerical error. This further proves the idea that the mismatch between the $\mu_{L}^{T}$ and $\bar{\mu}_{L}^{T}$ can be attributed to an artifact of the experiment. Finally, it should be noted that a similar mismatch between $\mu_{L}^{T}$ and $\bar{\mu}_{L}^{T}$ can occur for small values of the particle's Schmidt number ($Sc$, Eq.~\cref{eq:Sc}),~\cite{balboa13} which are typical for LB. However, as we have shown in Section~\ref{sub:dimnum} our particle Scmidt number is quite large ($Sc \ge 10^{3}$), which should put our system in the regime where the thermal and quiescent (deterministic) results correspond well.~\cite{balboa13}

\begin{figure}[!htb]
\begin{center}
\includegraphics[scale=1.0]{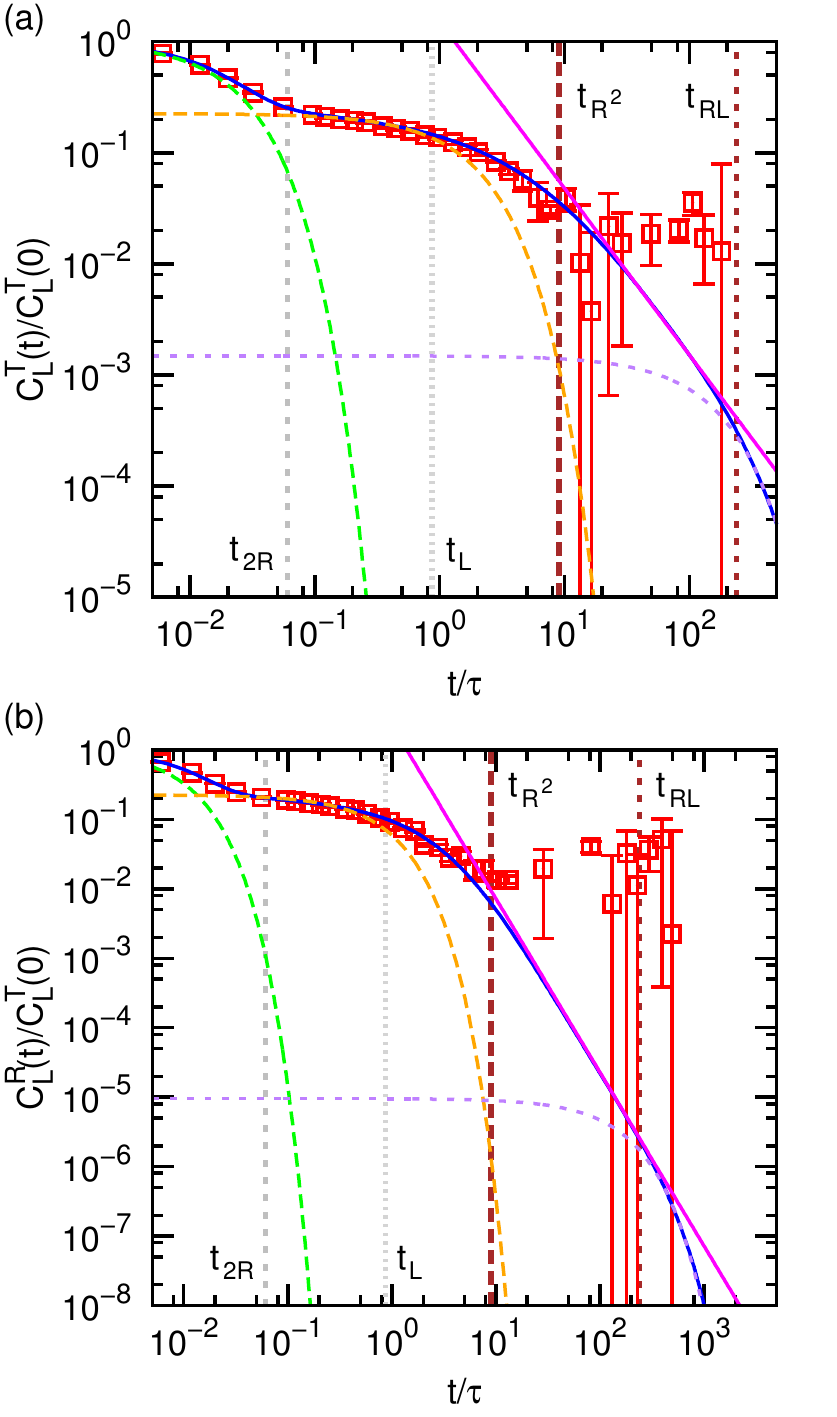}
\end{center}
\caption{\label{fig:vacf_hollow}(color online) Auto-correlation functions for a hollow raspberry of radius $R = 3.0\sigma$ in a box of length $L = 80.0\sigma$ as a function of the time $t$ expressed in the MD time unit $\tau$. The LB parameters are as given in the text. The use of symbols is the same as in Fig.~\ref{fig:vacf_filled}. (a) Velocity auto-correlation function (VACF) $C_{L}^{T}(t)$ for the thermal (red squares with error bars) and quiescent (blue curve) calculations. The unphysical coupling (dashed green), the fitted Stokes (orange dashed), fitted final exponential (purple dashed) curves, and power-law decay (magenta) are also shown. The two gray dashed vertical lines show the time it takes for sound to travel the particle's hydrodynamic diameter ($t_{2R}$, dashed) and the box length ($t_{L}$, dotted), respectively. The two brown dashed vertical lines show the time associated with viscous dissipation ($t_{R^{2}}$, dashed) and ($t_{RL}$, short dashes). (b) The angular-velocity auto-correlation function (AVACF) $C_{L}^{R}(t)$ for the same parameters.}
\end{figure}

$~$
\paragraph*{The Hollow Raspberry}
$~$\newline

In order to examine the difference between the hollow and filled raspberry model, we carried out similar experiments for a hollow-raspberry sphere in a box of length $L=80.0\sigma$. We find similar regimes as in Fig.~\ref{fig:vacf_filled}. For the hollow raspberry there is weaker coupling with the fluid. This is caused by the spatial distribution of coupling points and reduced the number of points. This results in weaker decay of the unphysical-coupling regime, which therefore matches the exponential form of Eq.~\cref{eq:VACFbal} more closely.

Note that the existence of the power-law behavior is more convincingly shown by our AVACF data, see Fig.~\ref{fig:vacf_hollow}(b), as the fitted function and measured decay correspond well over a decade in time. It is unclear whether the modified coupling scheme by Mackay~\textit{et al.}~\cite{Mackay13a} shows a similar decay. Finally, it should be noted that for the hollow raspberry VACF there is the same deviation between the quiescent and thermalized LB results as shown in Fig.~\ref{fig:vacf_filled}(a). However, the thermal and quiescent data for the AVACF match well throughout.

\subsubsection{\label{subb:FSS}The Influence of Crystal Lattice Spacing}

Thus far, we have examined only the results of the (angular) velocity experiments, and shown that these correspond -- at least amongst themselves -- to the results of force experiments for the same system. Let us now consider the effect of the lattice spacing of the simple cubic crystal on the hydrodynamic coupling between the spheres. This simple cubic geometry is unlike a physical crystal, in the sense that all particles translate and rotate in unison; an effect of there being only a single particle in a box with periodic boundary conditions. However, there is experimental evidence that such systems may be realized.~\cite{friese98,whitesides00,whitesides01} The uniformity of the periodic structure makes the solutions to Stokes' equations for this geometry analytically tractable. Such calculations were performed, for example, in the work of Hasimoto~\cite{hasimoto59} and of Hofman~\textit{et al.}~\cite{hofman99b}

\begin{figure*}
\begin{center}
\includegraphics[scale=1.0]{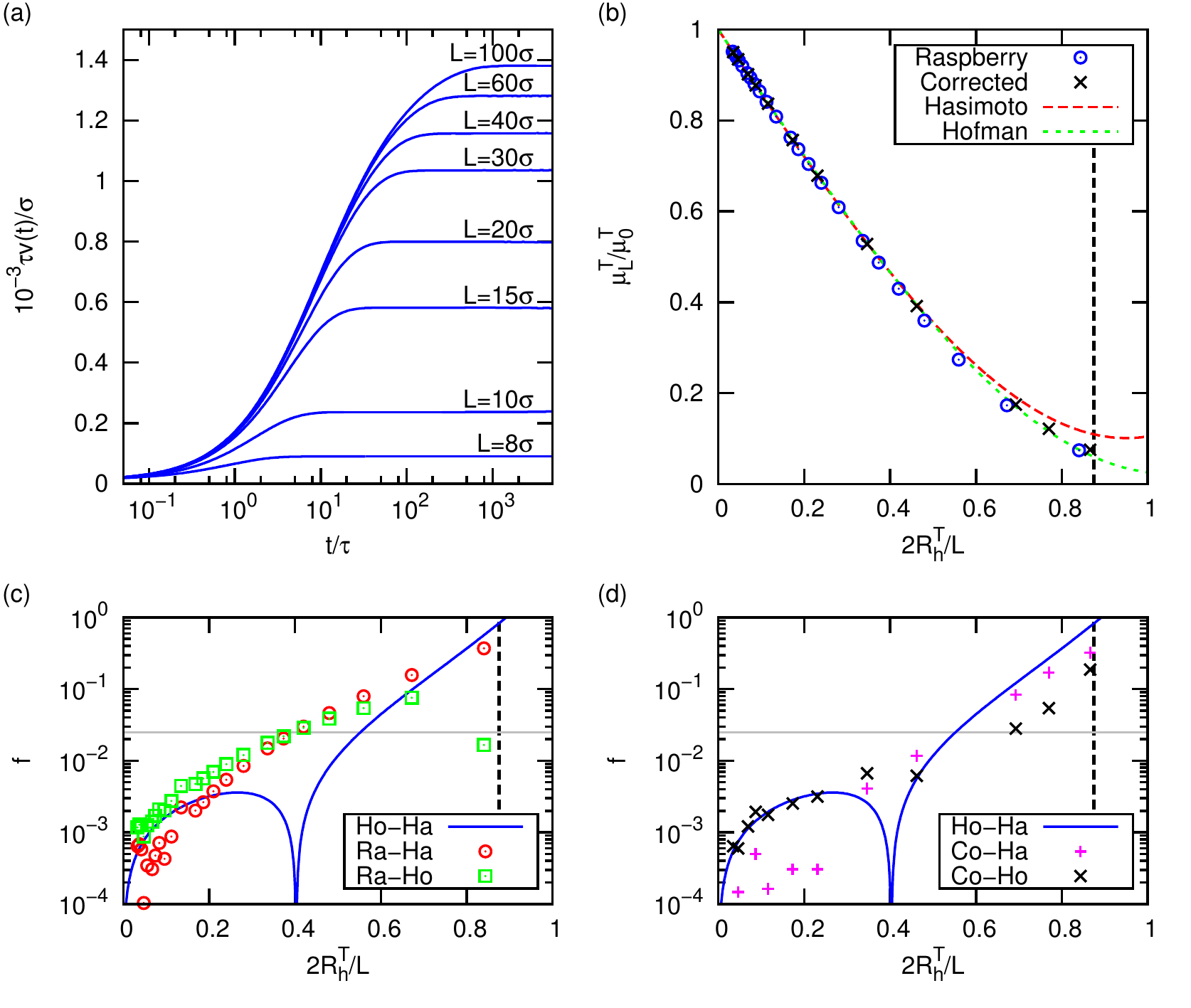}
\end{center}
\caption{\label{fig:tra_fss}(color online) (a) The velocity $v(t)$ as a function of the time $t$ expressed in the MD time unit $\tau$ obtained from force experiments for a selection of box sizes $L$. (b) The dependence of the particle's translational mobility $\mu^{T}_{L}$ (expressed in terms of the bulk translational mobility $\mu^{T}_{0}$) on the inverse box length $1/L$ times the hydrodynamic diameter (twice the hydrodynamic radius $R^{T}_{h}$). The blue circles show results obtained from the velocity experiment. The red dashed curve shows the analytic expression by Hasimoto~\cite{hasimoto59} (Eq.~\cref{eq:hasi}) and the green dashed curve shows a polynomial fit to the numerical data of Hofman~\textit{et al.},~\cite{hofman99b} see Eq.~\cref{eq:hofman_trans}. The black crosses give the results of the force experiment, corrected for the counter force applied inside of the raspberry, as explained in detail in the main text. The vertical dashed black line indicates the value of $L$ for which the spheres are separated by one lattice spacing $(L = 2R^{T}_{h} + \sigma)$. (c) Fractional deviation $f$ as a function of $2R^{T}_{h}/L$. The solid blue curve shows the difference between the theoretical expressions of Hasimoto and Hofman~\textit{et al.}. The red circles and green squares indicate the difference between our raspberry experiment and the Hasimoto and Hofman~\textit{et al.} expressions, respectively. (d) Fractional deviation $f$ for the force-corrected data. The blue line is the same as in (c). The magenta pluses and black crosses give the difference between the force-corrected data and the Hasimoto and Hofman~\textit{et al.} expressions, respectively. The gray horizontal line in (c,d) indicates a fractional deviation of 2.5\%.}
\end{figure*} 

$~$
\paragraph*{Translation of the Crystal}
$~$\newline

Figure~\ref{fig:tra_fss}(a) shows the change in velocity $v(t)$ during a quiescent force experiment (see Fig.~\ref{fig:cube}(a)) for a number of box sizes $L$ using the filled raspberry model. Note that for larger $L$ the friction experienced by the particle is smaller, as the hydrodynamic-interaction with its periodic images is reduced. However, the time it takes for the stationary state to set in is increased, as it takes longer to transfer momentum between the particle and its images. From the terminal velocities in the stationary state we determined the mobility, by averaging over several oscillations due to lattice artifacts. 

In order to establish the mobility at infinite dilution (one particle in bulk), we fitted our data using a polynomial of the Hasimoto form:~\cite{hasimoto59} $A + B/L + C/L^{3}$, see Eq.~\cref{eq:hasi}, in the range where this form is expected to be valid ($L \gtrsim 3 R$, as we will see later in this section) and extrapolated to $L\uparrow\infty$. The resulting value for $A$ is the bulk translational mobility
\begin{equation}
\label{eq:StEin} \mu^{T}_{0} \equiv \frac{1}{6 \pi \eta R^{T}_{h}},
\end{equation}
with $R^{T}_{h}$ the translational hydrodynamic radius. We were thus able to determine the extrapolated value $\mu^{T}_{0}$ and simultaneously the effective hydrodynamic radius of our raspberry colloid, using Eq.~\cref{eq:StEin}. This extrapolation refers to the `fitting' part of our `filling + fitting' formalism. These two parameters $\mu^{T}_{0}$ and $R^{T}_{h}$ allowed us to non-dimensionalize the box length and the measured translational mobility, as shown in Fig.~\ref{fig:tra_fss}b. 

In Figs.~\ref{fig:tra_fss}(b,c) we compare the quality of our result for the box-size dependence with the analytic result by Hasimoto~\cite{hasimoto59} given in Eq.~\cref{eq:hasi} (dashed red curve) and the numerical calculations by Hofman~\textit{et al.}~\cite{hofman99b} (dashed green curve). Figure~\ref{fig:tra_fss}(c) shows the fractional deviation $f$ between our data and the two literature results, as well as the difference between the Hasimoto (Ha) and Hofman~\textit{et al.} (Ho) data. For the data points provided by Hofman~\textit{et al.}~\cite{hofman99b} we used a polynomial fit to represent these as a curve. The fit has the following shape
\begin{equation}
\label{eq:hofman_trans} \mathscr{H}^{T}(R,L) = 1 - 2.807 \left( \frac{R}{L} \right)  + 3.437 \left( \frac{R}{L} \right)^{3} .
\end{equation}
Note that the analytic and numerical expressions of Refs.~\cite{hasimoto59,hofman99b} correspond well for box sizes greater than $L \approx 5.0 R^{T}_{h}$. That is, within the error expected for the fitting procedure that we applied to the data by Hofman~\textit{et al.}, there is good agreement between their results and Hasimoto's data over this range. The discrepancy for smaller box sizes can be explained by the truncation of the series expansion in Hasimoto's work.

Our raspberry results (Ra) agree reasonably well with the data of Hofman~\textit{et al.} over the range $L \gtrsim 5.0 R^{T}_{h}$, but there is also a clear signature of systematic deviation present in $f$. This implies that our data differs substantially from the values of Ref.~\cite{hofman99b} in the $1/L^{3}$ term. A similar range of agreement and small-box-size deviation can be observed between our data and that of Hasimoto. However, in spite of this, our data is much closer to the results of Hofman~\textit{et al.} than those of Hasimoto; by almost an order of magnitude in $f$ for $L \downarrow 2 R_{h}^{T}$. We will discuss the origin of this systematic deviation between our data and that of Ref.~\cite{hofman99b} next.

$~$
\paragraph*{Origin of the Discrepancy}
$~$\newline

The discrepancy between our data and the result by Hofman~\textit{et al.} brings us back to the difference that we observed between the VACFs obtained from the velocity and temperature experiments carried out in Section~\ref{subb:VACFsphere}, see Fig.~\ref{fig:vacf_filled}(a). Remember that in the quiescent experiments a homogeneous and instantaneous velocity has to be applied to the fluid in order to ensure zero net movement of the system, see Fig.~\ref{fig:cube}(e). Similarly, for the quiescent force experiment, a constant homogeneous force density is applied to the fluid, see Fig.~\ref{fig:cube}(a). Consequently, this velocity and force are also applied directly to the fluid nodes that are coupled to the raspberry MD beads. The effective force applied to the colloid can therefore be calculated by subtracting the integrated fluid force-field over the volume of the raspberry. This calculation yields
\begin{equation}
\label{eq:Feff} \mathbf{f}_{\mathrm{eff}} = \mathbf{f}\left( 1 - \frac{4 \pi R^{3}}{3 L^{3}} \right),
\end{equation}
where $\mathbf{f}$ is the force directly applied to the central bead of our raspberry construct. Analogously, the counter velocity affects the time evolution of the VACF. For the thermalized experiments this was not an issue, since counter velocities and forces do not need to be applied. These counter velocities and forces are therefore a likely candidate for the observed discrepancies. This implies that the force/velocity experiments are unsuited to analyze the hydrodynamic properties of finite systems in their present form. The fact that there is a mismatch between the thermal and quiescent results in Fig.~\ref{fig:vacf_filled}(a) is thus \underline{not} an expression of a violation of the equipartition theorem or fluctuation dissipation. \underline{Nor} is it correct to argue that this is a consequence of the porosity of the particle. The counter-force is only used to counter momentum transfer to the periodic system by the force applied to the particle. The behavior in the limit of the infinite system is, however, accurately captured, as the back velocity and force vanish.

We took the effective force of Eq.~\cref{eq:Feff} to determine the `corrected' value of $\mu^{T}_{L}$ (Co) using Eq.~\cref{eq:ES} as a function of the box size, see Fig.~\ref{fig:tra_fss}(b). Note that the correspondence between the result by Hofman~\textit{et al.} and our data is thus greatly improved and that the systematic deviation is removed for large box sizes, see Fig.~\ref{fig:tra_fss}(d). Moreover, for small box sizes the deviation between our corrected result and the literature values is substantially reduced, although a systematic difference remains. Within the error, the data corresponds much closer to the data by Hofman~\textit{et al.} than it does to the Hasimoto result.

From our corrected data, we estimated the range over which the raspberry is able to accurately reproduce hydrodynamics interactions ($f < 2.5\%$) in our system. For this particular model we found the criterion to be $L \gtrsim 2.8 R^{T}_{h}$, which can be extrapolated to other spatial arrangements of the colloids. It is likely that this criterion can be extended to smaller boxes, as we will see in the following and in Part II.~\cite{degraaf15}. The normalized results for a hollow raspberry lie on top of the filled ones shown in Fig.~\ref{fig:tra_fss}(b) within the error bar (not shown here). However, the values for the effective hydrodynamic radii $R^{T}_{h}$ differ: $3.53\sigma$ and $3.47\sigma$ for the filled and hollow model, respectively. 

\begin{figure}[!htb]
\begin{center}
\includegraphics[scale=1.0]{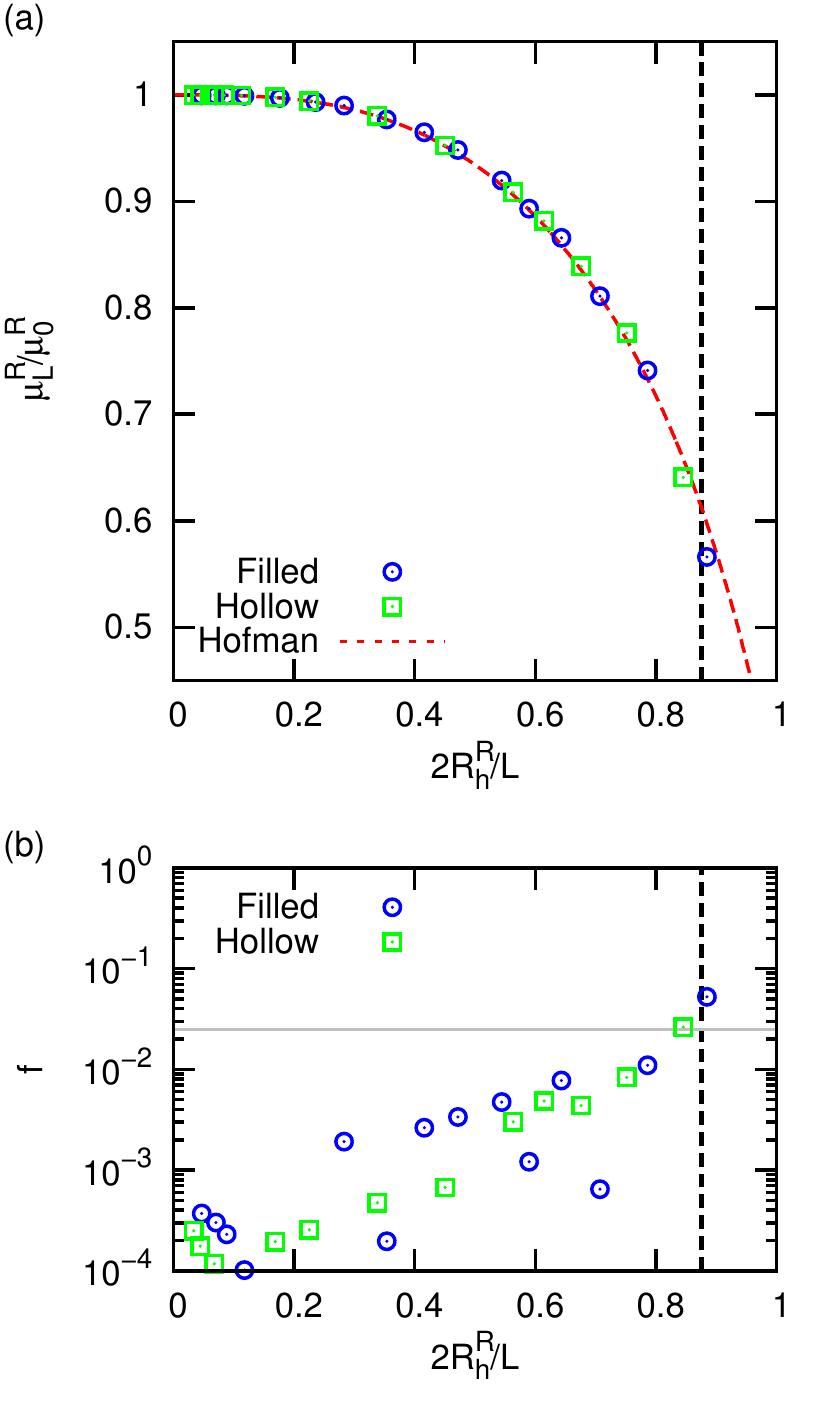}
\end{center}
\caption{\label{fig:rot_fss}(color online) (a) The dependence of the particle's rotational mobility $\mu^{R}_{L}$ (expressed in terms of the bulk rotational mobility $\mu^{R}_{0}$) on the inverse box length times the hydrodynamic diameter (twice the hydrodynamic radius $R^{R}_{h}$). The blue circles show results obtained for the filled raspberry and the green squares for the hollow raspberry. The red dashed curve shows the expression given by Hofman~\textit{et al.},~\cite{hofman99b} see Eq.~\cref{eq:hofman_rot}. The vertical dashed black line indicates the value of $L$, for which the spheres are separated by one lattice spacing. (b) Fractional deviation $f$ as a function of $2R^{R}_{h}/L$. The blue circles and green squares indicate the difference between the filled and hollow raspberry and analytic expression, respectively. The gray horizontal line indicates a fractional deviation of 2.5\%.}
\end{figure}

$~$
\paragraph*{Rotation in the Crystal}
$~$\newline

We continued our verification of the quality of the filled and hollow raspberry model by examining hydrodynamic coupling between spheres rotating in unison in a cubic lattice, as before, see Fig.~\ref{fig:cube}(c). Figure~\ref{fig:rot_fss} shows a comparison of our results to the expression given by Hofman~\textit{et al.}~\cite{hofman99b} for the box-size dependence of the rotational mobility $\mu^{R}_{L}$. The expression provided in Ref.~\cite{hofman99b} reads
\begin{equation}
\label{eq:hofman_rot} \mathscr{H}^{R}(R,L) = 1 - 4.189 \left( \frac{R}{L} \right)^{3}  + 231.858 \left( \frac{R}{L} \right)^{10} .
\end{equation}
The procedure used to generate this data is analogous to that outlined for the translational experiments. Using
\begin{equation}
\label{eq:StEinR} \mu^{R}_{0} \equiv \frac{1}{8 \pi \eta \left(R^{T}_{h}\right)^{3}}
\end{equation}
we determined the effective hydrodynamic radius $R^{R}_{h}$ from our data. Note that while there is still a systematic component to $f$, see Fig.~\ref{fig:rot_fss}(b), the agreement between our result and literature is excellent for both models. 

This further demonstrates the plausibility of our assertion that the high level of deviation for the translational mobility is caused by the back-force/velocity that is applied homogeneously to the fluid, since a similar correction is not required for the rotational experiments. However, there is a fundamental difference between the experiments. The rotational motion exposes the fluid to constantly varying coupling points (the MD beads), whereas for translational motion the fluid could more easily find a pathway of least resistance. This could be another source of discrepancies in the translational experiments not present in the rotational experiments.

We again observed that the effective hydrodynamic radii obtained for the hollow and filled raspberry differ significantly, $3.38\sigma$ and $3.54\sigma$, respectively. It should be stressed that the fact that behavior of $\mu^{R}_{L}$ is the same for both models, does not imply hydrodynamic consistency of the model, when we compare the value of $R^{T}_{h}$ and $R^{R}_{h}$ for the same model, which we will do next.

\begin{figure}[!htb]
\begin{center}
\includegraphics[scale=1.0]{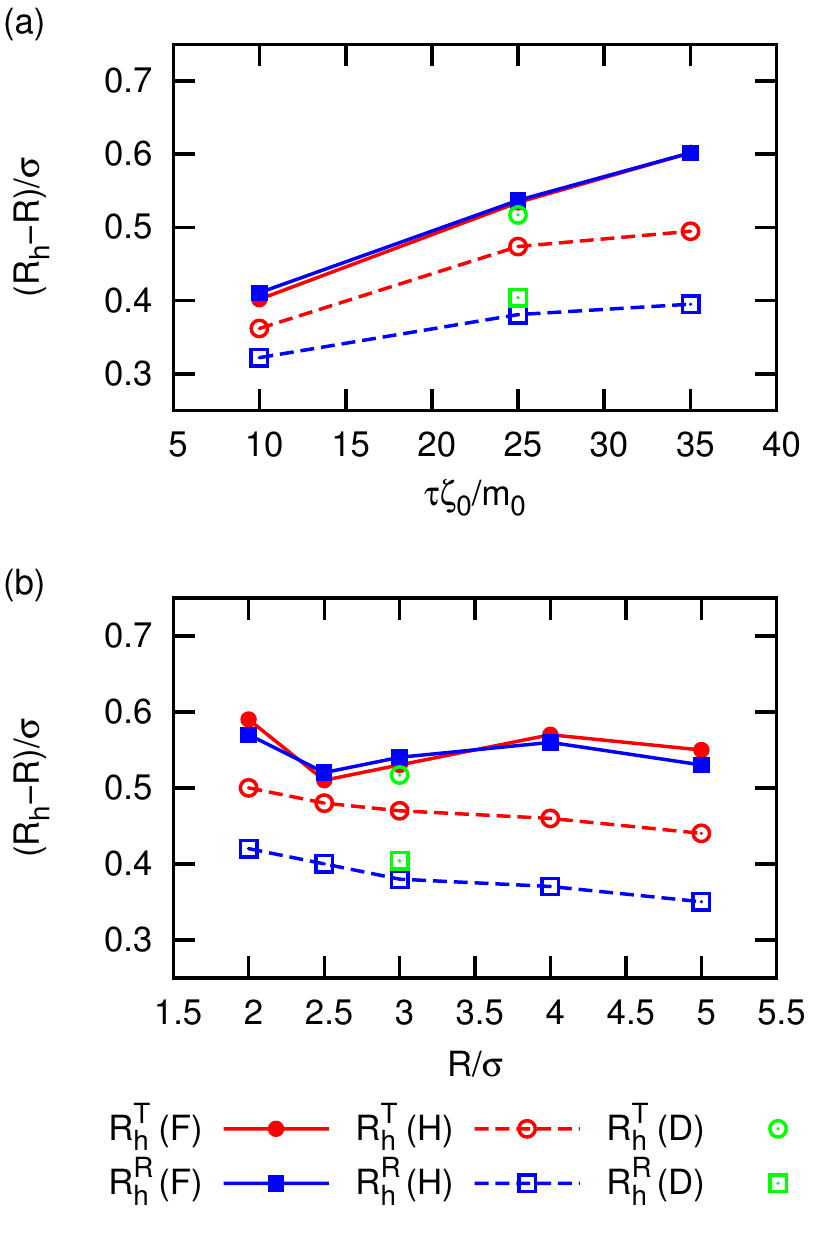}
\end{center}
\caption{\label{fig:Reff}(color online) The difference of the bulk effective hydrodynamic radii (translational $R^{T}_{h}$, rotational $R^{R}_{h}$) with respect to the imposed radius $R$ for the filled (F), hollow (H), and dense shell (D) raspberry models, respectively. (a) The difference as a function of the bare friction coefficient  $\zeta_{0}$ expressed in LB units (time $\tau$ and mass $m_{0}$) for raspberry particles with a radius of $R = 3.0\sigma$. (b) The difference as a function of the imposed radius $R$ for a bare particle-fluid friction of $\zeta_{0} = 25 m_{0}\tau^{-1}$.}
\end{figure}

\subsubsection{\label{subb:porous}The Effective Radius}

$~$
\paragraph*{The Radius Dependence on Various Parameters}
$~$\newline

To further assess the significance of the difference between the effective hydrodynamic radii, we repeated our experiments for two other values of the bare friction $\zeta_{0}$ and several $R$. The results for the box-size dependence were in quantitative agreement. Our results for the hydrodynamic radii are summarized in Fig.~\ref{fig:Reff}. These were obtained, as before, by extrapolating to the bulk value of the mobility. The translational and rotational radii of the hollow raspberry differ substantially. This result is in agreement with the findings of Ollila~\textit{et al.}~\cite{Ollila12,ollila13} and it is in line with the theoretical predictions of Refs.~\cite{Debye48,Felderhof75a,Felderhof75b} The mismatch occurs for all values of the friction coefficient and radius that we examined. This discrepancy is, however, undesirable to simulate hard colloidal spheres, a purpose for which the raspberry model was initially introduced.~\cite{lobaskin04} In each case, the agreement between the effective hydrodynamic radii of the filled raspberry particles is almost perfect.

We also performed experiments using a hollow raspberry with $N_{\mathrm{tot}} = 925$ -- the same total number of beads as in the filled raspberry -- we refer to this model as the `dense shell' raspberry. This allowed us to examine the hypothesis that we simply obtained an increased effective friction with the greater bead numbers used in the filled raspberry, leading to a better match between rotational and translational hydrodynamic radius.~\cite{ollila13} Also note that the screening ratio for the filled and dense raspberry is the same, see Table~\ref{tab:kappa}. A discrepancy between $R^{T}_{h}$ and $R^{R}_{h}$ was found for the dense shell raspberry, see Fig.~\ref{fig:Reff}. In fact, the deviation is slightly larger than for the $N = 202$ hollow raspberry. This can be attributed to an overall improvement of the coupling in the dense shell raspberry, which forces the translational radius towards the no-slip value more quickly than the rotational one.

\begin{figure}[!htb]
\begin{center}
\includegraphics[scale=1.0]{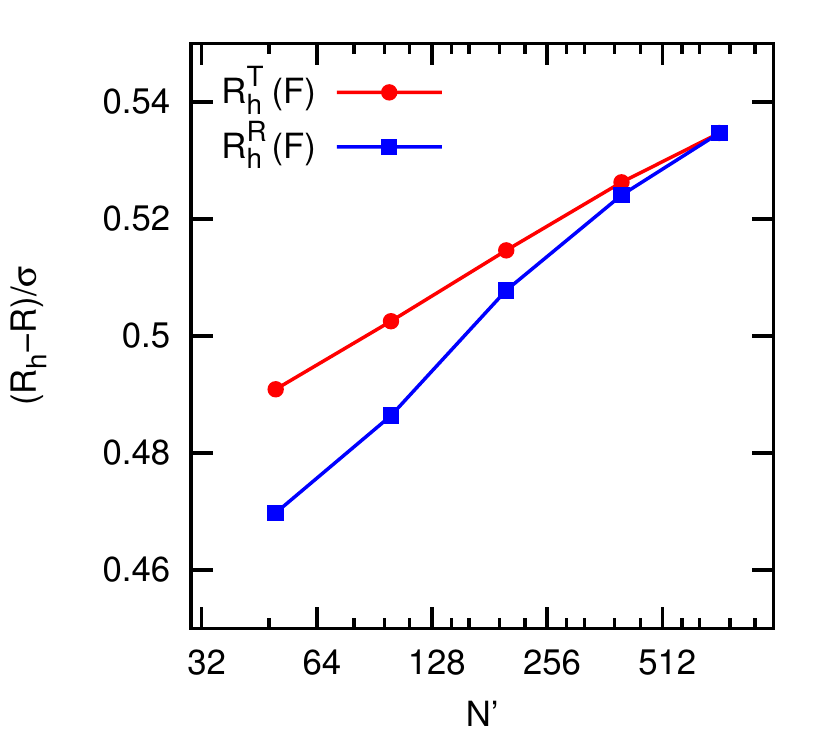}
\end{center}
\caption{\label{fig:fill}(color online) The difference of the bulk effective hydrodynamic radii as a function of the number of internal coupling points $N'$ for a raspberry with radius $R = 3.0\sigma$. The dots and connecting red curve show the translational hydrodynamic radius $R^{T}_{h}$ and the blue squares with connecting curve show the rotational hydrodynamic radius $R^{R}_{h}$.}
\end{figure}

To investigate the impact of the level of filling on the particle, we varied the number of internal coupling points $N'$ for a raspberry with radius $R = 3.0\sigma$ with $N = 202$ shell-coupling points. The result is shown in Fig.~\ref{fig:fill}, which gives the dependence of the hydrodynamic radii on the filling parameter $N'$. It is clear that the correspondence between $R^{T}_{h}$ and $R^{R}_{h}$ can be substantially improved by adding coupling points, until there is essentially no longer a difference, at our chosen value of $N' = 722$. This correspondence is reached at a feasible number of coupling points. However, it requires considerably more than one coupling point per lattice cell, \textit{i.e.}, a filling density greater than $10.0 a^{-3}$ was found to give almost perfect correspondence between the two radii.

\begin{figure}[!htb]
\begin{center}
\includegraphics[scale=1.0]{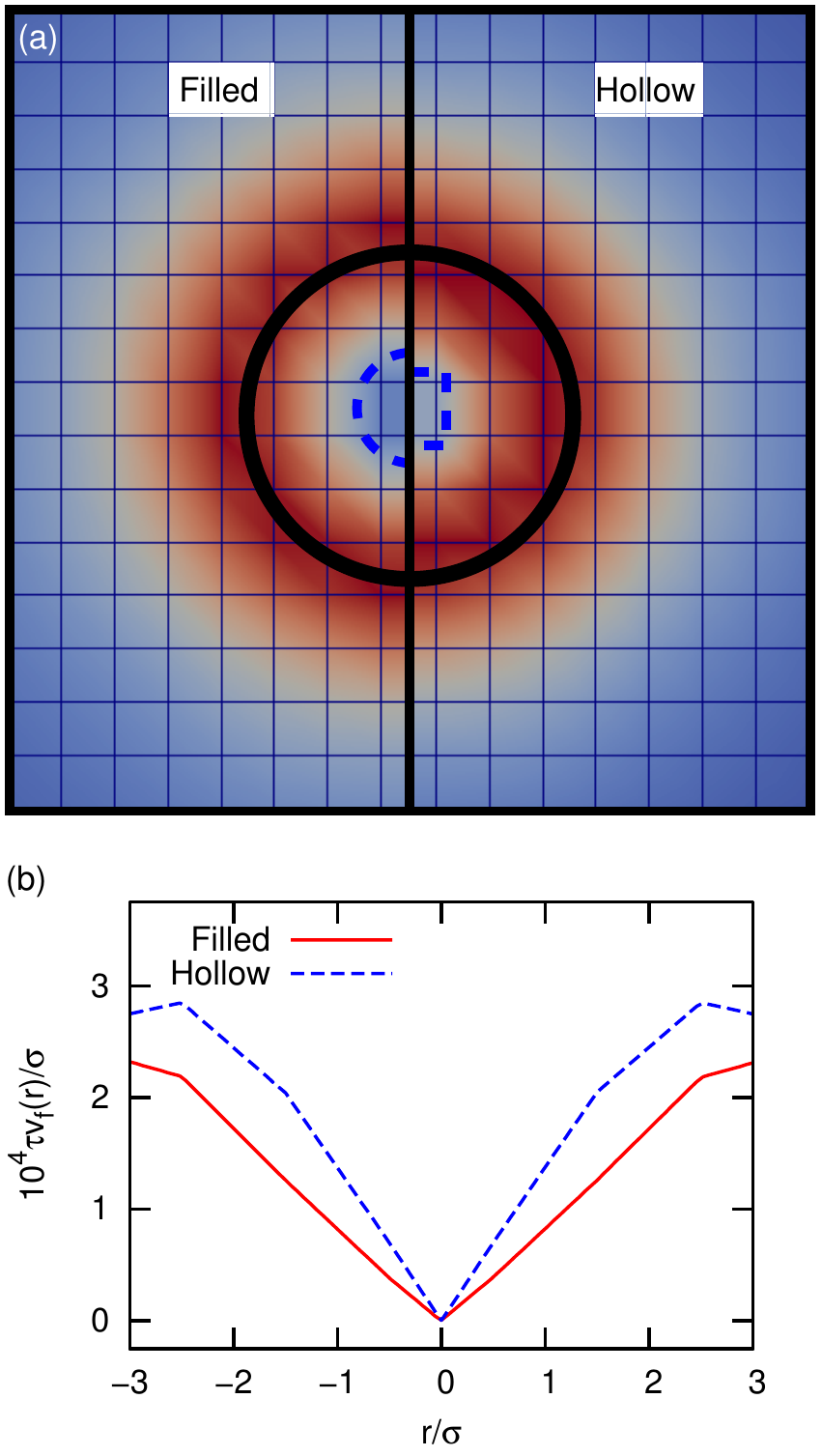}
\end{center}
\caption{\label{fig:compare}(color online) Comparison of the flow field around a filled and hollow raspberry, respectively, undergoing a constant rotation. (a) Two dimensional plane through the center of the sphere with a normal that is parallel to the axis of rotation. The result for the filled raspberry is shown on the left and for the hollow variant on the right. The color coding gives the magnitude of the fluid velocity on the grid (blue lines). The thick black circle roughly indicates the position of the coupling points (at $\vert \mathbf{r} \vert = R$). The dashed blue semi circle and half square serve as guides to the eye for the structure of the flow field inside the raspberry. (b) Magnitude of the fluid velocity $v_{f}(r)$ -- expressed in MD units of time, $\tau$, and position $\sigma$ -- as a function of position $r$ along the black vertical divide in (a). Only the value inside of the raspberry is shown for the filled (red, solid) and hollow (blue, dashed) particle.}
\end{figure}

We examined the fluid flow inside the filled and hollow $R = 3.0\sigma$ raspberry with $N_{\mathrm{tot}} = 925$ and $203$ respectively, to determine the cause of the inconsistency between the effective hydrodynamic radii for the hollow model. Figure~\ref{fig:compare}(a) shows the flow field around a hollow and filled spherical raspberry, rotating at constant angular velocity about the axis pointing into the page. From the flow field it becomes apparent that the coupling of the raspberry to the fluid has more lattice artifacts (is less smooth) for the hollow raspberry than for the filled one, indicating poorer coupling. We quantified this difference further by examining the fluid velocity inside the particle, see Fig.~\ref{fig:compare}(b). While the filled raspberry shows a linear increase in the velocity with the distance from the center (similar to the so-called `Rankine vortex state'), the hollow raspberry shows a clear kink in the velocity profile. This kink can be attributed to the diminished fluid-particle coupling away from the shell of MD beads. Effectively, the hollow shell raspberry achieves its Screening ratio (a low Brinkman length) only close to the shell, whereas the filled raspberry achieves low permeability throughout.

\begin{figure}[!htb]
\begin{center}
\includegraphics[scale=1.0]{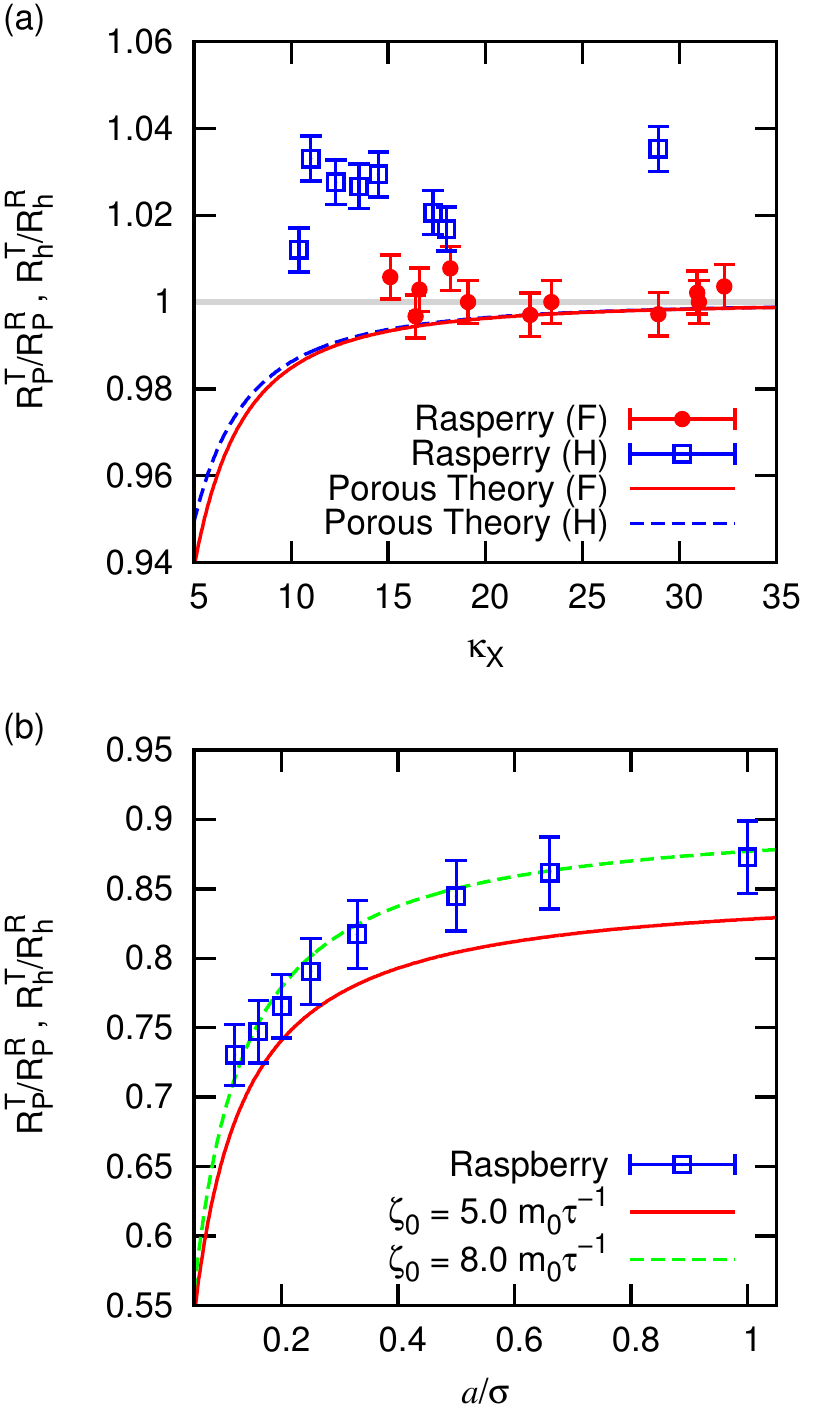}
\end{center}
\caption{\label{fig:porous}(color online) Comparison to the results of theory for porous spheres.~\cite{Debye48,Felderhof75a,Felderhof75b} (a) The ratio of the translational and rotational hydrodynamic radii as a function of the screening ratio $\kappa_{\mathrm{X}}$ (`X' is either `F' for filled or `H' for hollow). The ratio $R^{T}_{\mathrm{P}}/R^{R}_{\mathrm{P}}$ from the theory~\cite{Debye48,Felderhof75a,Felderhof75b} is indicated by a red solid curve for a filled sphere and a blue dashed curve is for a hollow shell. The ratio $R^{T}_{h}/R^{R}_{h}$ for the raspberry simulations are indicated using symbols with standard error bars; red dots show the filled raspberry results and blue open squares those for the hollow raspberry. The thick gray line indicates the unit ratio. (b) The ratios as a function of the lattice spacing $a$ that were obtained for a hollow raspberry with radius $R = 3.0\sigma$, $N = 25$ coupling points on the surface, and a bare friction of $\zeta_{0} = 5.0 m_{0}\tau^{-1}$. The blue squares with error bars show the results of our raspberry simulations. The red solid curve shows the prediction of the theory for this system. The dashed green line shows the prediction of the theory for a slightly higher bare friction ($\zeta_{0} = 8.0 m_{0}\tau^{-1}$).}
\end{figure}

$~$
\paragraph*{The Porosity of the Raspberry}
$~$\newline

Finally, let us discuss our results in the context of the predictions made by theory for porous spheres.~\cite{Debye48,Felderhof75a,Felderhof75b} In Fig.~\ref{fig:porous}(a) we have plotted the theoretical prediction for the ratio of the hydrodynamic radii $R^{T}_{\mathrm{P}}/R^{R}_{\mathrm{P}}$ as a function of the screening ratio $\kappa_{\mathrm{X}}$ (`X' is either `F' for filled or `H' for hollow). This data is based on Table~\ref{tab:kappa} and Eqs.~(\ref{eq:srfill}-\ref{eq:HporRR}). We also show the ratio $R^{T}_{h}/R^{R}_{h}$ that we obtained from our raspberry simulations. It is clear that there is a mismatch between our results and the predictions of theory. That is to say, the trends predicted by theory are not reproduced. This can be attributed to the fact that the theory solves Stokes' equation with a Stokeslet point-coupling. The reality of the finite grid-size LB simulations is that the point-coupling only approximates the Stokeslet.~\cite{ahlrichs99} Correspondence is only found at a few lattice spacings away from the coupling point and the Stokeslet form can only be reproduced in the limit of small lattice spacings $a$.~\cite{dunweg09} From Fig.~\ref{fig:porous}(a) it becomes clear that for finite $a$, the translational hydrodynamic radius is larger than that of the rotational one; the opposite of the theoretical prediction.~\cite{Debye48,Felderhof75a,Felderhof75b}

This leads to the question: ``Can the result of the theory in principle be reproduced by our simulations?'' In order to determine this, we chose parameters which are unsuited to achieve our goal of obtaining hydrodynamic correspondence, but allow for a sufficiently low fluid-particle coupling to observe the difference in radii. For a hollow raspberry with radius $R = 3.0\sigma$, $N = 25$ coupling points on the surface, and a bare friction of $\zeta_{0} = 5.0 m_{0}\tau^{-1}$, the theory predicts a radius ratio of less than one and a strong decay of this ratio with the lattice spacing, see Fig.~\ref{fig:porous}(b). The curve is based on a combination of Eqs.~(\ref{eq:srfill}-\ref{eq:HporRR}). Since the computation time scales as $a^{-3}$, we used a fixed box size $L = 20\sigma$. This allowed us to use grid spaces of $a = 0.125\sigma$, i.e., LB grids with $160^{3}$ elements, which is roughly the limit of the grid size that fits into a modern GPU's memory. We therefore did \underline{not} perform finite-size scaling. We exploited the Hasimoto relation~\cite{hasimoto59} of Eq.~\cref{eq:hasi} to fit for the effective translational hydrodynamic radius and the Hofman~\textit{et al.}~\cite{hofman99b} relation of Eq.~\cref{eq:hofman_rot} to fit for the effective rotational hydrodynamic radius. 

Figure~\ref{fig:porous}(b) shows the result of our simulations. The error bars are sizable, but appropriate for the limited box-sizes that we could study upon varying $a$. It is clear that for these parameters our data has the same trend as predicted by the theory. However, we found that a slightly higher bare friction coefficient, namely $\zeta_{0} = 8.0 m_{0}\tau^{-1}$, yields better agreement with our simulation result. This difference can be attributed to the fact that the theory assumes distribution of the screening ratio $\kappa_{\mathrm{H}}$ that is homogeneous over the shell, whereas our numerical results are for individual coupling points. For a higher number of coupling points, the decay in $R^{T}_{\mathrm{P}}/R^{R}_{\mathrm{P}}$ is very weak for reasonable LB parameters, which makes it difficult to see if the theoretically predicted trends are matched within the error bar. An additional source of discrepancy is the effective shell-width of $\sim 2a$.~\cite{dunweg09} At the maximum resolution that we were able to achieve, there is still an effective width of $0.25\sigma$, whereas the theory assumes a dirac-delta distributed Screening ratio for the hollow raspberry.~\cite{Debye48,Felderhof75a,Felderhof75b} Considering these two sources of error, the agreement with theory that we were able to achieve is quite excellent. With sufficient computational resources, the porosity prediction should be captured for a more dense distribution of shell coupling points and even smaller value of $a$, but this falls outside of the scope of our current investigation.

\subsection{\label{sub:SCdumb}Dumbbell in a Simple Cubic Crystal}

Thus far, we have concentrated on the quality of the raspberry approximation for convex objects, namely the specific case of a spherical particle. In order to assess the raspberry model's ability to capture the hydrodynamic properties of a non-convex particle, we considered two dumbbell-shaped raspberries, as shown in Fig.~\ref{fig:dumb}. We took care to create a dumbbell raspberry model for which the two spheres touch, when the effective hydrodynamic radius of the MD beads is taken into account, see Fig.~\ref{fig:dumb} (left). Note that for a dumbbell-shaped particle the hydrodynamic mobility tensor (HMT) has a diagonal form, with translational mobilities in the top-left $3\times3$ block (sub-matrix) and rotational ones in the lower-right $3\times3$ block. There are no cross-coupling terms due to symmetry.~\cite{Brenner65,Brenner67}

Our results for the dumbbell particles are qualitatively similar to those shown for the spherical colloid discussed above. Namely, we found the box-size dependence to be of the form $\mu^{X}_{L,i} = \mu_{0,i}^{X} \left( 1 + A_{i}/L + B_{i}/L^{3} \right)$, with $X$ either $R$ or $T$ and $i$ either $\perp$ or $\parallel$, and $A_{i}$ and $B_{i}$ coefficients. However, we could not compare our results to analytic calculations, since, to the best of our knowledge, such expressions have not been formulated for dumbbell-shaped particles. We therefore considered the extrapolated bulk mobility coefficients only. Using both quiescent and thermalized simulations we verified that the HMT has the expected form. In particular, all off-diagonal coefficients were orders of magnitude smaller than the diagonal elements and zero within the error bars. Moreover, we found that for both the translational and rotational mobility sub-matrices, the two entries corresponding to perpendicular motion were equal (within the error) and the parallel component was larger, as expected. Table~\ref{tab:compare} lists these mobility coefficients. In order to non-dimensionalize the results, we divided the mobility coefficients by the translational and rotational mobility of a sphere with radius $R = 3.0\sigma$ (the size of one of the dumbbell's lobes), respectively.

To validate our model for the simulation of anisotropic non-convex particles, we compared our data with the results obtained using the \textit{HYDROSUB} and \textit{HYDRO++} program.~\cite{garcia81,garcia07} These are tools used to evaluate the hydrodynamic properties of macromolecules and have been successfully utilized in comparisons to experimental data for solid anisotropic colloids.~\cite{kraft13} We determined the HMT using the methods of Refs.~\cite{garcia81,garcia07} for dumbbells consisting of two spheres with radii $R = 1.0$ $\mu$m at positions $(\pm1.0,0,0)$ $\mu$m (touching) and $(\pm0.714,0,0)$ $\mu$m (overlapping), respectively, in a fluid of viscosity $1.0\cdot 10^{-3}$ kg$\,$m$^{-1}\,$s$^{-1}$ and density $1.0\cdot 10^{3}$ kg$\,$m$^{-3}$ with temperature $T = 293.15$ K. We assumed that the particle has the same density as the fluid. The numerical algorithm is parametrized as follows: $H = 26$, $H_{\max} = 1.5\cdot10^{7}$, $R_{\max} = 80.0\cdot10^{-8}$, and $N_{\mathrm{TRIALS}}$ = 10,000; which are internal commands. The number of intervals for the distance distribution was set to 30. By applying the same numerical parameters to the case of a single sphere we obtained the reference data used to normalize the result, which in turn allows for a direct comparison to our results.

\begin{table}
\begin{ruledtabular}
\begin{tabular}{c|c|c|c|c}
Method & $\mu^{T}_{\parallel}/\mu^{T}_{0}$ & $\mu^{T}_{\perp}/\mu^{T}_{0}$ & $\mu^{R}_{\parallel}/\mu^{R}_{0}$ & $\mu^{R}_{\perp}/\mu^{R}_{0}$ \\
\hline
\multicolumn{5}{c}{$d = 7 \sigma$ / $d = 2.00$ $\mu$m} \\
\hline
Rasp. (H)                & $0.78\pm0.01$ & $0.70\pm0.01$ & $0.61\pm0.01$ & $0.28\pm0.01$ \\
Rasp. (F)                & $0.77\pm0.01$ & $0.69\pm0.01$ & $0.55\pm0.01$ & $0.27\pm0.01$ \\
\cite{garcia07} & $0.77\pm0.01$ & $0.70\pm0.01$ & $0.55\pm0.01$ & $0.27\pm0.01$ \\
\hline
\multicolumn{5}{c}{$d = 5 \sigma$ / $d = 1.43$ $\mu$m} \\
\hline
Rasp. (H)                & $0.83\pm0.01$ & $0.75\pm0.01$ & $0.67\pm0.01$ & $0.39\pm0.01$ \\
Rasp. (F)                & $0.82\pm0.01$ & $0.74\pm0.01$ & $0.59\pm0.01$ & $0.36\pm0.01$ \\
\cite{garcia81} & $0.82\pm0.01$ & $0.75\pm0.01$ & $0.60\pm0.01$ & $0.37\pm0.01$ \\
\end{tabular}
\end{ruledtabular}
\caption{\label{tab:compare}Comparison for a dumbbell-shaped particle between the results obtained using the raspberry model -- both hollow (H) and filled (F) -- and \textit{HYDROSUB}/\textit{HYDRO++}~\cite{garcia81,garcia07} for the translational (T) and rotational (R) mobilities in the direction parallel ($\parallel$) and perpendicular ($\perp$) to the main axis in bulk fluid. The mobilities are normalized by the bulk values for a sphere with the same radius as one of the spheres comprising the dumbbell.}
\end{table}

The results of this comparison are summarized in Table~\ref{tab:compare}, in which we give the mobilities for the filled and hollow raspberry, as well as the ones determined using the methods of Refs.~\cite{garcia81,garcia07}. The agreement for the translational bulk mobilities is excellent in all three data sets. However, it is clear that for the hollow raspberry there is a significant difference in the rotational mobility ratio with respect to the result for the filled and \textit{HYDROSUB}/\textit{HYDRO++} simulations. This difference lies well outside of the error bar of the average of the latter two. This confirms that our `filling + fitting' procedure is effective for more complex (non-convex) geometries, as expected.

\section{\label{sec:disc}Discussion}

In Section~\ref{sec:result} we have demonstrated that our `filling + fitting' formalism leads to excellent agreement between established theoretical and numerical results for the hydrodynamic behavior of convex and non-convex solid particles. By `filling + fitting' one significantly improves the agreement between the effective hydrodynamic radii obtained by translational and rotational experiments, respectively, allowing the point-coupling LB model to describe solid particles. The improvement is related to a reduced permeability throughout the particle -- in line with the findings of Ref.~\cite{Ollila12}. The hollow-shell raspberry achieves this only locally.~\cite{lobaskin04,chatterji05,ollila13,Mackay13b} In this section we discuss this discrepancy between the effective hydrodynamic radii in more detail and place our work in the context of previous studies.

The fractional difference in hydrodynamic radii of approximately $0.1\sigma/3.4\sigma = 0.03$ for the hollow raspberry may seem perfectly acceptable for most applications. However, one should be careful, since this small fraction can lead to a 10\% discrepancy between the expected translational and rotational mobility, had we assumed the effective hydrodynamic radius for rotational motion to be the same as that for translational motion. In processes involving both translation and rotation, this could lead to significant deviation from the desired behavior. 

$~$
\paragraph*{Previous Studies}
$~$\newline

A closer examination of the data presented in the original raspberry paper by Lobaskin and D{\"u}nweg~\cite{lobaskin04} shows that the trends in matching to the results of Refs.~\cite{hasimoto59,zick82,brenner70,zuzovsky83,hofman99b} with effective hydrodynamic radii observed in our work, are captured by their data points. Lobaskin and D{\"u}nweg erroneously assumed that the radius of the particle was the same as the radius $R$ at which they positioned their MD beads. Within the numerical uncertainty present in their results and the computational abilities of the time, this extrapolation to bulk was unavoidable. By re-examining the data points of Ref.~\cite{lobaskin04}, we conclude that it is possible to fit the following bulk mobilities
\begin{eqnarray}
\label{eq:DTlob} \mu^{T}_{0} & = & (0.97\pm0.02)\frac{k_{\mathrm{B}}T}{6 \pi \eta R}; \\
\label{eq:DRlob} \mu^{R}_{0} & = & (0.90\pm0.02)\frac{k_{\mathrm{B}}T}{8 \pi \eta R^{3}}.
\end{eqnarray}
This indicates that there is indeed an effective radius, $R_{h}^{T} = (1.03 \pm 0.02) R$ and $R_{h}^{R} = (1.03 \pm 0.01)R$, but the data is not of sufficient quality to assess whether there is a difference between the effective translational and rotational hydrodynamic radius in their measurements.

Chatterji and Horbach~\cite{chatterji05} carried out a more thorough examination of the effective translational hydrodynamic radius. However, they did not provide results for the rotational hydrodynamic radius, they only comment on having carried out such experiments. Our results in Fig.~\ref{fig:Reff} for the value of $R^{T}_{h}$ for the hollow raspberry are in quantitative agreement with Ref.~\cite{chatterji05}. We therefore deem it likely that a similar discrepancy would be present in the data of Ref.~\cite{chatterji05}, especially considering our observations and those of Refs.~\cite{Ollila12,ollila13}.

Finally, Poblete~\textit{et al.}~\cite{poblete14} did not report a difference in the bulk hydrodynamic radii using their MPCD method for a hollow raspberry. They instead found agreement between the two. However, it is unclear how accurately Poblete~\textit{et al.} could extrapolate their results to the bulk value, as in MPCD one always works with thermalized and therefore noisy data. In addition, it is not obvious how large the effect ($R^{T}_{h} \ne R^{R}_{h})$ would be for their high-speed-of-sound systems. Furthermore, the grid-shifts that are typically applied in MPCD to restore Galilean invariance, may substantially reduce any such lattice-discretization and porosity effects.

$~$
\paragraph*{Relation to the Work of Ollila et al.}
$~$\newline

The inconsistency between the translational and rotational mobility in the raspberry model was first pointed out by Ollila~\textit{et al.}~\cite{Ollila12,ollila13} Reference~\cite{Ollila12} contends that these inconsistencies are representative of the properties of the point-coupling method. Namely, that the objects modeled using this formalism are porous. Ollila~\textit{et al.} argue that this porosity leads to problems when using this type of model to describe solid objects. In particular, models that fit for the radii should be considered with suspicion according to Ref.~\cite{Ollila12}, as the fitted hydrodynamic radii may be inconsistent between various hydrodynamic experiments. This assessment may seem in direct contradiction to our observations. However, Ollila~\textit{et al.} do not exclude the possibility of finding numerical parameters for which a quality fit can be made. We have shown here, as well as in Ref.~\cite{degraaf15}, that our `filling + fitting' formalism works well to match the simulations to analytic results for solid particles over a wide range of parameters. That is, we obtained numerically consistency for physically relevant hydrodynamic experiments. 

Note that the excellent agreement shown between the simulation results and analytic expressions for porous spheres in Ref.~\cite{Ollila12} is not without caveats. In particular, Ollila~\textit{et al.} indicate that it is necessary to use a particle radius for the coupling points that is `incommensurate' with the lattice to obtain the excellent correspondence for the translational properties of the porous particles without fitting. Due to the properties of the interpolation scheme, this incommensurability criterion and the subsequent choice of a particle radius that yields correspondence, can be treated on the same footing as a fit parameter. Moreover, Ollila~\textit{et al.} require an effective hydrodynamic radius (another fit) to obtain similar correspondence for the rotational properties of their particles.

We have performed our simulations with both stationary and moving particles at positions and in directions both commensurate and incommensurate with the lattice. In all these experiments, we did not find a sizable change in the effective radii, nor a breakdown of the correspondence between the two. We thus argue that our `filling + fitting' method is a cleaner and more forthright way of proving a correspondence between a theoretical result and simulations. We therefore believe that an equally excellent correspondence between theory and simulations could have been achieved in Ref.~\cite{Ollila12}, by dropping the incommensurability criterion and fitting for both effective hydrodynamic radii. In addition, our `filling + fitting' method is an excellent approach to find LB and coupling parameters for which the behavior of solid objects in a Stokes' fluid can be faithfully reproduced.

$~$
\paragraph*{Numerical Efficiency Considerations}
$~$\newline

In both Refs.~\cite{Ollila12,ollila13} the number of coupling points used to obtain correspondence between theory and simulations is rather large. Such a high number of points is acceptable in addressing questions of a fundamental nature, but it may prove problematic in performing simulations with many raspberry particles, as is typical for self-assembly and crystallization studies.~\cite{roehm14}

The algorithm may become prohibitively expensive for these numbers of coupling points. Lowering the overall number of coupling points and specifically their local density is of particular relevance to GPU-based LB implementations. The force applied to the nodes of the LB grid by a coupling point is calculated using so-called `atomicAdd' operations. These operations can be used to avoid race conditions that arise from colliding memory requests. However, for a large number of beads close to a specific LB node (high coupling-point density), the use of the atomicAdd operation can cause the program to slow down significantly. Therefore, reducing the number of local coupling points is of paramount importance and our `filling + fitting' procedure is thus numerically favorable to models that require a higher local coupling-point density.

$~$
\paragraph*{Porosity and the Filling Factor}
$~$\newline

With regards to the filling procedure, we obtained reasonable consistency for the translational and rotational mobility for roughly 10 beads per LB grid volume $a^{3}$ within the particle, for a grid size of $a = 1.0\sigma$. Depending on the radius of the particle, we obtained a slight variation of the effective increase in hydrodynamic size of the particle, but for all of our $R$ and $\zeta_{0}$ data points, we managed to achieve far superior agreement between the hydrodynamic radii for the filled raspberry than for the hollow variant. 

As mentioned above, in discussing the results by Ollila~\textit{et al.}~\cite{Ollila12,ollila13}, our results seem counterintuitive, when one considers theoretical predictions for porous media.~\cite{Debye48,Felderhof75a,Felderhof75b} Indeed, we find that for the LB parameters that we chose (which are physically reasonable), our results deviate substantially from the predicted trends. However, one should be careful in interpreting these results. The LB method solves a discretized form of the Boltzmann transport equation, which only reproduces the result of Stokes' equation in the appropriate limits (small grid spacing, etc.). This means that point-forces that are applied to the LB fluid via the Ahlrichs and D{\"u}nweg interpolated point-coupling scheme~\cite{ahlrichs99} are not true Stokeslets. Therefore, it is a priori not to be expected that the theoretical result is reproduced. We have demonstrated that the results of porous sphere theory can in principle be recovered in our simulations, when the surface coverage of coupling points and the bare friction coefficient are sufficiently small. Whether this result can be extended to greater particle numbers is difficult to assess with the precision we can currently achieve.

$~$
\paragraph*{The Short-Time Behavior}
$~$\newline

We end with a comment on the short-time behavior of the raspberry particles. As originally shown in Ref.~\cite{lobaskin04}, the Ahlrichs and D{\"u}nweg interpolated point-coupling scheme~\cite{ahlrichs99} has problems in reproducing the short-time properties of the (A)VACF that are expected for a solid no-slip particle~\cite{Zwanzig75} or even a porous colloid,~\cite{Felderhof14} due to the presence of an unphysical coupling regime. The correct zero-time value of $C^{T}_{L}(0) = 3 k_{\mathrm{B}}T/m$ is achieved, but there is no decay to $C^{T}_{L}(t>0) = 3 k_{\mathrm{B}}T/m^{\ast}$, with $m^{\ast}$ the virtual mass, over a time scale related to the propagation of sound.~\cite{Zwanzig75} Instead, a much lower plateau value for $C^{T}_{L}(t>0)$ is reached. Felderhof~\cite{Felderhof14} has pointed out that the secondary (virtual mass) regime is \underline{not} present for a porous colloid in the high frequency limit. It is possible that our findings are in agreement with this observation and that the decay with $m^{\ast}$ reported by Lobaskin and D{\"u}nweg.~\cite{lobaskin04} is not to be expected. However, the presence of the unphysical coupling regime and the difficulty in determining the limit in which our LB is operating frustrate further analysis at this time. An analysis of the influence of both the frequency and strength of the viscous coupling on the result and the correspondence to the predictions of Ref.~\cite{Felderhof14} is left for future study.

It has been suggested that the modified coupling scheme by Mackay~\textit{et al.}~\cite{Mackay13a} may resolve the short-time decay issues. However, examination of the VACFs reported in Ref.~\cite{Mackay13b} reveal that the double exponential-type decay shown in their fluid-mass dominated result is not captured by the result of Ref.~\cite{Zwanzig75}, as is reported in Ref.~\cite{Mackay13b} (but not demonstrated using fitting procedures). This is, again, expected on the basis of the results by Felderhof.~\cite{Felderhof14} Unfortunately, it is also not clear that the predictions of Ref.~\cite{Felderhof14} are more accurately reproduced. Furthermore, an analysis of the results of the oscillatory experiments in Ref.~\cite{Ollila12} does not yield significant additional insight into the short-time quality of the Mackay~\textit{et al.} algorithm. In particular, it is unclear whether the Mackay~\textit{et al.} coupling algorithm was used in Ref.~\cite{Ollila12}. In addition, the period of the oscillation is sufficiently long to obfuscate any short-time discrepancies that may be present.

\section{\label{sec:conc}Conclusion and Outlook}

Summarizing, we have re-examined the properties of the hybrid LB and Langevin MD scheme for simulating colloids developed by Lobaskin and D{\"u}nweg,~\cite{lobaskin04} the so-called `raspberry' model. We studied this model using a variety of classic fluid dynamics experiments that predominantly focused on the long-time mobility properties of these particles. We considered the hydrodynamic properties of spherical raspberries, as well as dumbbell-shaped raspberry particles in the low Reynolds number limit. Our results show that the proper solid-particle mobility in this limit is reproduced to a surprising degree of accuracy over a wide range of viscosities for both convex and non-convex particle shapes.

From our combined data we can draw the following conclusions concerning the quality of the raspberry model and our `filling + fitting' procedure to match its hydrodynamic properties to that of a solid object in a Stokes' fluid.
\begin{itemize}
\item Using a raspberry model to approximate a particle's coupling to an LB fluid gives rise to an effective hydrodynamic radius. Our result is in agreement with the findings of Chatterji and Horbach.~\cite{chatterji05} 
\item The short-time properties of a no-slip or permeable colloid are not faithfully reproduced by the Ahlrichs and D{\"u}nweg interpolated point-coupling scheme,~\cite{ahlrichs99} as was first pointed out in Ref.~\cite{lobaskin04}.
\item The traditional `hollow' raspberry model -- an empty shell of MD coupling beads that describes the particle's surface -- gives rise to a discrepancy between the translational and rotational effective hydrodynamic radius. This effect was first pointed out by Ollila~\textit{et al.}~\cite{Ollila12,ollila13} and discussed in the context of porous particle dynamics.~\cite{Debye48,Felderhof75a,Felderhof75b}
\item We find that the aforementioned mismatch, when considered in the context of reproducing the hydrodynamic behavior of a solid particle, can be reduced to within an acceptable numerical tolerance by `filling' the raspberry and `fitting' for the effective hydrodynamic radius. Here, we found a filling density of about 10 coupling points per LB grid cell gives satisfactory results.
\item The `filling + fitting' procedure is not in disagreement with the assessment of Ollila~\textit{et al.}~\cite{Ollila12} that such a filling procedure is inherently problematic. Our result demonstrates that for reasonable LB parameters the hydrodynamic properties of a solid particle can be effectively matched and to within a far higher tolerance than is possible for the hollow variant. 
\item Moreover, the formalism is not in contradiction with the theoretical results for porous particles.~\cite{Debye48,Felderhof75a,Felderhof75b} The fact that we can obtain excellent correspondence between the hydrodynamic radii for a filled raspberry and we find inferior agreement for the hollow variant, should be considered in the context of the level of discretization that is used. Hence, a mismatch between the theoretical predictions and the numerical result is not unexpected; in this case it can be exploited.
\item The `filling + fitting' procedure can be used to improve the raspberry model's ability to simulate both convex and non-convex solid particles. We verified this for the specific case of a dumbbell-shaped particle, and our procedures may be safely extrapolated to more complicated shapes. 
\item The result of Ollila~\textit{et al.}~\cite{ollila13} suggest that a regime can be found for which the hydrodynamic hull is sufficiently shrunk that it matches with the imposed position of the coupling points. However, a prohibitive number of coupling points may be required to achieve this condition. This is especially problematic for GPU-based algorithms. Our `filling + fitting' procedure allows us to use a substantially reduced number of coupling points and still obtain excellent numerical agreement.
\item The force and velocity experiments traditionally performed to determine the translational mobility in a cubic geometry with periodic boundary conditions are problematic for small boxes compared to the particle size. The back force/velocity density that must be applied to the fluid to maintain zero center of mass velocity, leads to difficulties in interpreting the mobility data that is obtained from these experiments. A possible solution to this problem is to identify the node locations at which the particle is found and to only apply the properly-rescaled counter force elsewhere.
\end{itemize}

From the above, it becomes clear that the raspberry model is an excellent way to approximate the long-time regime of the fluid-particle coupling for a solid object in an LB fluid. However, there remains several open problems to be addressed in future studies. We have shown that the short-time behavior of the raspberry model (for the LB parameters used in this manuscript) is quite different from the low Reynolds number solution to Stokes' equations.~\cite{Zwanzig75,Felderhof14} This raises the question of how accurately the short-time regime of colloid dynamics can be captured using the raspberry or any point-coupling model. A faithful reproduction of such short-time processes would be relevant for, e.g., nucleation and crystallization.~\cite{roehm14} Despite this concern, our analysis stresses the power of our `filling + fitting' method as a means to find parameters for which the translational as well as rotational hydrodynamic properties of the original raspberry model can be sufficiently enhanced to give rise to Stokesian fluid - solid particle coupling.

\section*{\label{sec:ack}Acknowledgements}

J.d.G. acknowledges financial support by a ``Nederlandse Organisatie voor Wetenschappelijk Onderzoek'' (NWO) Rubicon Grant (\#680501210). We thank the ``Deutsche Forschungsgemeinschaft'' (DFG) for financial funding through the SPP 1726 ``Microswimmers -- from single particle motion to collective behavior''. We are also grateful to O. Hickey, U. Schiller, and S. Kesselheim for useful discussions.

\end{document}